\documentclass[9pt,twocolumn,twoside]{osajnl}

\journal{ao} 

\usepackage[caption=false]{subfig}

\setboolean{shortarticle}{false} 

\title{Field-only integral equation method for time domain scattering of electromagnetic pulses}

\author[1]{Evert Klaseboer}
\author[2]{Qiang Sun}
\author[3,4*]{Derek Y. C. Chan}

\affil[1]{Institute of High Performance Computing, 1 Fusionopolis Way, Singapore 138632, Singapore}
\affil[2]{Particulate Fluids Processing Center, Department of Chemical Engineering, The University of Melbourne, Parkville VIC, 3010, Australia}
\affil[3]{Particulate Fluids Processing Center, School of Mathematics and Statistics, The University of Melbourne, Parkville VIC, 3010, Australia}
\affil[4]{Department of Mathematics, Swinburne University of Technology, Hawthorn VIC, 3122, Australia}

\affil[*]{Corresponding author: D.Chan@unimelb.edu.au}

\ociscodes{(290.5825) Scattering theory, (290.4020) Mie theory, (290.4210) Multiple scattering.}

\begin{abstract}
The scattering of electromagnetic pulses is described using a non-singular boundary integral method to solve directly for the field components in the frequency domain, and Fourier transform is then used to obtain the complete space-time behavior. This approach is stable for wavelengths both small and large relative to characteristic length scales. Amplitudes and phases of field values can be obtained accurately on or near material boundaries. Local field enhancement effects due to multiple scattering  of interest to applications in microphotonics are demonstrated.
\end{abstract}

\setboolean{displaycopyright}{true}

\begin{document}
\maketitle
\thispagestyle{fancy}

\ifthenelse{\boolean{shortarticle}}{\ifthenelse{\boolean{singlecolumn}}{\abscontentformatted}{\abscontent}}{}

\section{Introduction}

Current theoretical formulations and computational algorithms for time domain electromagnetic problems are well-developed for applications that range from transient spectroscopy~\cite{Meyer2016} to telemetry involving high speed spacecrafts~\cite{Garner2017}.
Under conditions relevant to micro-photonics in which the interest in regimes of wavelengths can be small or large compared to the characteristic dimensions of the scatterers or the need to have accurate values of the phases and amplitudes of the field near boundaries, current methodologies encounter challenges.

This paper addresses the time domain electromagnetic problem of the scattering of electromagnetic pulses by working directly in terms of the components of the electric field using a non-singular surface integral formulation in the frequency domain. 
Fourier transform is then used to give the complete space-time behavior. 

This approach retains the advantages of reduction in spatial dimension of surface integral methods, and because of the non-singular nature of the surface integrals, it has the added ability to handle surface geometric intricacies that often arise in micro-photonics problems in which certain length scales in the problem many be small compared to the wavelength or where there is the need to obtain accurate results for field values on or near surfaces. 
The use of the Fourier transform to give the time evolution also avoids possible instabilities that can arise with march on time algorithms.
Examples of the space-time dependence of scattering of an incident wave pulse by conducting and dielectric scatterers are provided to illustrate the transition between wave and geometric optics. 
The structure of the field near the surfaces is used to demonstrate field focusing and multiple scattering effects.

\section{Background}

One established approach to time domain electromagnetics is based on extending the surface integral formulations, the electric field integral equation (EFIE) or magnetic field integral equation (MFIE) method, to the time domain using a march on time method, see for example,~\cite{Jung2003}, ~\cite{Jung2004} and~\cite{Jung2010} for reviews. 
This widely used approach involves solving surface integral equations for the surface currents.
The electric and magnetic field are then obtained by post-processing the surface current values.
In present formulations, the surface integral equations that need to be solved for the induced surface current densities contain singular kernels that originate from the Green's function. This mathematical feature that does not have a physical basis means a loss of precision in the calculation of field values close to boundaries between different media. 
Furthermore, march on time algorithms can also lose precision as time progresses.

Another approach to time domain electromagnetics is the finite difference time domain (FDTD) method of Yee~\cite{Yee1966} where the space- and time-dependent Maxwell's equations are discretized in the 3D space variables and time stepping is used to track the space-time evolution from given initial conditions.
Although the FDTD algorithm is simple conceptually, there are a number of technical issues that require care in implementation. 
Convergence constraints impose limitations on the step sizes in time and space. 
Numerical dispersion effects associated with the relative orientation of the spatial grid and the direction of propagation can arise and unphysical reflections can occur at the boundaries (that conform to the stepwise nature of the grid) between regions of the different grid densities.
If the problem domain is infinite, an outer boundary needs to be constructed with suitable boundary conditions to satisfy the radiation condition at infinity so as to avoid unphysical reflections back into the solution domain~\cite{Taflove1988}.

Earlier works on pulses have been based on extending the analytic Mie theory for scattering by a sphere~\cite{Shifrin1994,Kim1996} and as such are not readily applicable to consider problems involving general scatterers.

\begin{figure}[t]     
\centering
    \includegraphics[width=0.3\textwidth]{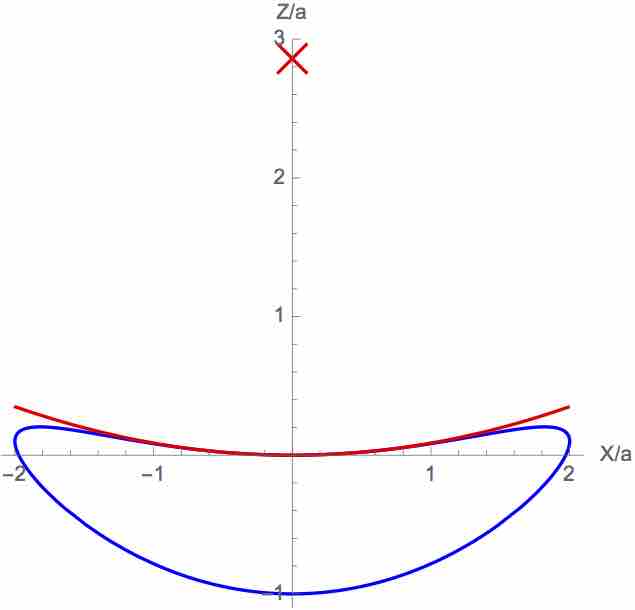}          
 \caption{The parametric curve (blue) in Eq.~\ref{eq:bowl_eqn} with $\beta = 0.6$ and $\gamma = 0.5$ together with the parabola $Z = X^2/(4f)$ (red) that is fitted to the concave inner surface of the bowl. The focus is marked at $(X,Z) = (0,f) = (0,2a/(2\beta - \gamma) = (0, 2.83)$.}
\label{fig:SI_bowl-curve}
\end{figure}

\section{Theory}
The present theoretical development is motivated by the Mie theory~\cite{Liou1977} in which the solutions of Maxwell's equations in the frequency domain are constructed from 2 scalar Debye potentials that obey the Helmholtz equation. 
Although the Debye potentials can be found by solving boundary integral equations, they are not suitable for numerical implementation because the boundary conditions on the electric, $\boldsymbol{E}$ and magnetic, $\boldsymbol{H}$ fields involve the first and second derivatives of these scalar potentials.

The following is a summary of the field only, non-singular boundary integral formulation of the solution of Maxwell's equation in the frequency domain. Detailed development and numerical implementation issues have been given in the literature~\cite{Klaseboer2017a, Sun2017}.

For a vector field that satisfies wave equation
\begin{equation}
  \nabla^2 \boldsymbol{E} + k^2 \boldsymbol{E} = \boldsymbol{0}
\label{eq:Helm_E}
\end{equation}
the divergence free condition on $\boldsymbol{E}$ can be written as
\begin{equation}
 2 \nabla \cdot \boldsymbol{E} \equiv \nabla^2 (\boldsymbol{x} \cdot \boldsymbol{E}) + k^2 (\boldsymbol{x} \cdot \boldsymbol{E}) = 0
\label{eq:Helm_xE}
\end{equation}
whereby the scalar function $(\boldsymbol{x} \cdot \boldsymbol{E})$, where $\boldsymbol{x}$ is the position vector, also satisfies the scalar wave equation or the Helmholtz equation.

Therefore the solution of Maxwell's equation for the scattered electric field, $\boldsymbol{E}$ can be expressed as the simultaneous solution of 4 coupled scalar Helmholtz equations of the general form
\begin{eqnarray} \label{eq:Helm_p}
  \nabla^2 p_i(\boldsymbol{x}) + k^2 p_i(\boldsymbol{x}) = 0,  \qquad i = 1 .. 4
 \end{eqnarray} 
where $p_i(\boldsymbol{x})$ denotes $(\boldsymbol{x} \cdot \boldsymbol{E})$ or one of the 3 Cartesian components of $\boldsymbol{E}$.
These 4 equations are coupled by the usual boundary conditions on the normal and tangential components of $\boldsymbol{E}$ at material boundaries.
The scalar function $(\boldsymbol{x} \cdot \boldsymbol{E})$ is the result of the application of the angular momentum operator on one of the Debye potentials~\cite{Low2004}. 

Writing the solution of Eq.~\ref{eq:Helm_E} as surface integral equations furnishes 3 relations between the 6 unknowns: $E_{\alpha}$ and $\partial E_{\alpha} / \partial n$, $(\alpha =x,y,z)$, where $\partial/\partial n \equiv \boldsymbol{n} \cdot \nabla$ and $ \boldsymbol{n}$ is the outward unit normal of the surface, $S$ of the solution domain. The boundary integral solution of Eq.~\ref{eq:Helm_xE} for the quantity $(\boldsymbol{x} \cdot \boldsymbol{E})$ provides one more relation between $E_{\alpha}$ and $\partial E_{\alpha} / \partial n$ since: $\partial (\boldsymbol{x} \cdot \boldsymbol{E}) / \partial n = \boldsymbol{n} \cdot \boldsymbol{E} + \boldsymbol{x} \cdot \partial \boldsymbol{E} / \partial n$. The electromagnetic boundary conditions on the continuity of the tangential components of $\boldsymbol{E}$ provide the remaining 2 equations to determine $\boldsymbol{E}$ and $\partial \boldsymbol{E} / \partial n$ completely. At the surface of a perfect electrical conductor, the tangential component of the incident field must be canceled by that of the scattered field.

The boundary integral solution of Eq.~\ref{eq:Helm_p} for the scattered field is based on Green's Second Identity that gives a relation between $p_i(\boldsymbol{x})$ and its normal derivative $\partial{p_i}/\partial{n}$ at points $\boldsymbol{x}$ and $\boldsymbol{x}_0$ on the boundary, $S$.  All singularities associated with the Green's function $G \equiv G(\boldsymbol{x},\boldsymbol{x}_0) = \exp(ikr) / r$, as $r \equiv |\boldsymbol{x} - \boldsymbol{x}_0|\rightarrow 0$ can be removed analytically by using the following non-singular formulation of the boundary integral equations~\cite{Klaseboer2017a,Sun2017}
\begin{eqnarray} \label{eq:NS_BIE}
\int_{S}^{} {[p_i (\boldsymbol{x}) - p_i (\boldsymbol{x}_0) g(\boldsymbol{x}) - \frac{\partial {p_i(\boldsymbol{x}_0)}} {{\partial {n}}}  f(\boldsymbol{x})] \frac{\partial {G}} {{\partial {n}}} dS(\boldsymbol{x}}) = \qquad \nonumber \\  \int_{S}^{} {G [ \frac{\partial {p_i (\boldsymbol{x})}} {{\partial {n}}} - p_i (\boldsymbol{x}_0) \frac{\partial {g (\boldsymbol{x})}} {{\partial {n}}} - \frac{\partial {p_i(\boldsymbol{x}_0)}} {{\partial {n}}} \frac{\partial {f (\boldsymbol{x})}} {{\partial {n}}} ]dS(\boldsymbol{x}}). 
\end{eqnarray} 
The requirement on the functions $f(\boldsymbol{x})$ and $g(\boldsymbol{x})$ is that they satisfy the Helmholtz equation and the following conditions at the point $\boldsymbol{x} = \boldsymbol{x}_0$ on surface, $S$:~\cite{Klaseboer2017a,Sun2017}. 
\begin{subequations}
\begin{align}
f(\boldsymbol{x}_0) & = 0 \\
\boldsymbol{n} \cdot  \nabla f(\boldsymbol{x}_0) & = 1 \\
g(\boldsymbol{x}_0) & = 1 \\
\boldsymbol{n} \cdot \nabla g(\boldsymbol{x}_0) & = 0
\end{align}
\label{eq:f_and_g}
\end{subequations}
Therefore, if $p_i$ (or $\partial{p_i}/\partial{n}$) is given, then Eq.~\ref{eq:NS_BIE} can be solved for $\partial{p_i}/\partial{n}$ (or $p_i$) in a straightforward manner. The reason is that for $f(\boldsymbol{x})$ and $g(\boldsymbol{x})$ that obey the above conditions, Eq.~\ref{eq:f_and_g}, the terms that multiply $G$ and $\partial G/\partial n$ vanish at the same rate as the rate of divergence of $G$ or $\partial G/\partial n$ as $\boldsymbol{x} \rightarrow \boldsymbol{x}_0$ and consequently both integrals have non-singular integrands and can thus be evaluated accurately by quadrature, see~\cite{Klaseboer2017a,Sun2017} for details. 
Consequently, this approach affords higher numerical precision with fewer degrees of freedom and confers numerically robustness that enables, in particular, the accurate calculation of field values on or near boundaries without adverse numerical issues.
Note that the solid angle at $\boldsymbol{x}_0$ that occurs in the traditional boundary integral equation has also been eliminated in Eq.~\ref{eq:NS_BIE}. Suitable choices for $f(\boldsymbol{x})$ and $g(\boldsymbol{x})$ can be found in~\cite{Klaseboer2017a,Sun2017}.
Similar coupled equations also hold for the magnetic field $\boldsymbol{H}$, see~\cite{Klaseboer2017a, Sun2017} for details. 
 
Furthermore, the present field only formulation is not affected by the so called \emph{zero frequency catastrophe} as $k \rightarrow 0$ that limits the accuracy of the familiar electric and magnetic field integral equations for the surface current density~\cite{Vico2016}.

As to be expected, the system of equations to be solved are simpler for perfect electrical conductor (PEC) scatterers~\cite{Klaseboer2017a} than for dielectric scatterers~\cite{Sun2017}, but the underlying physical concepts are the same. 

\begin{figure}[t]     
\centering
  \begin{tabular}{@{}cc@{}}
    \includegraphics[width=.23\textwidth]{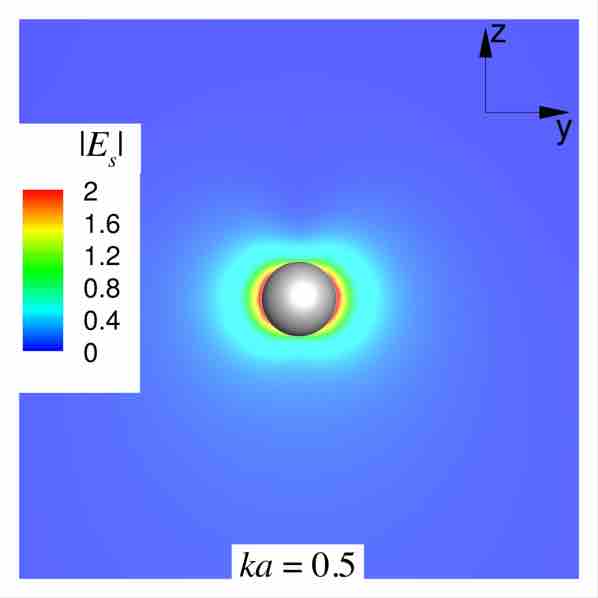} &
    \includegraphics[width=.23\textwidth]{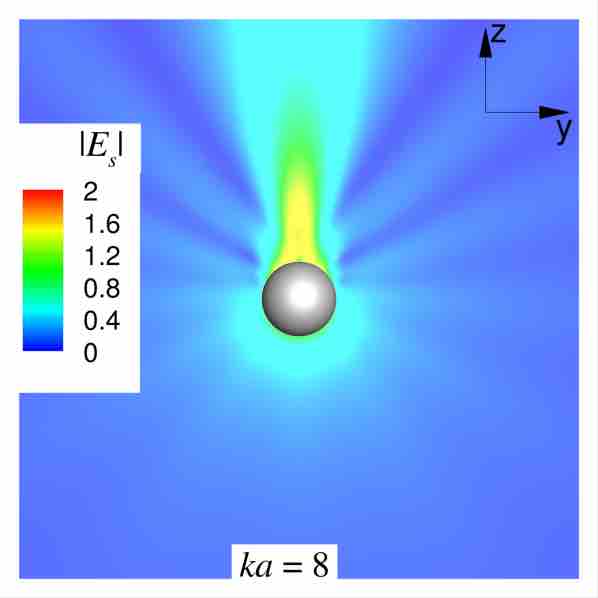}\\ 
    \includegraphics[width=.23\textwidth]{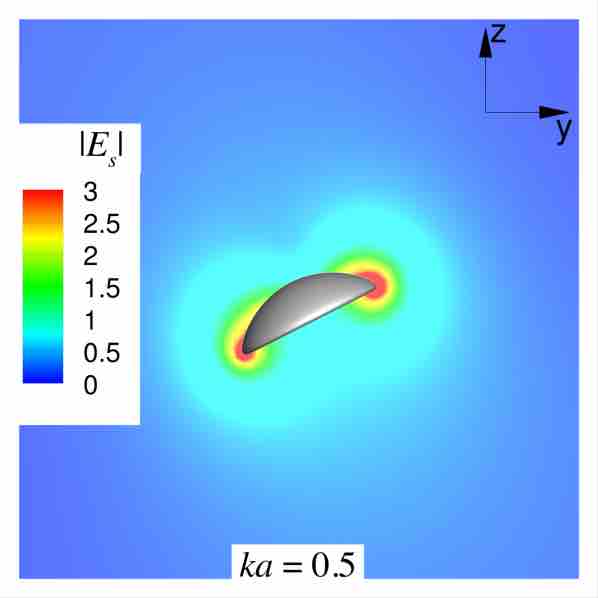}  &
    \includegraphics[width=.23\textwidth]{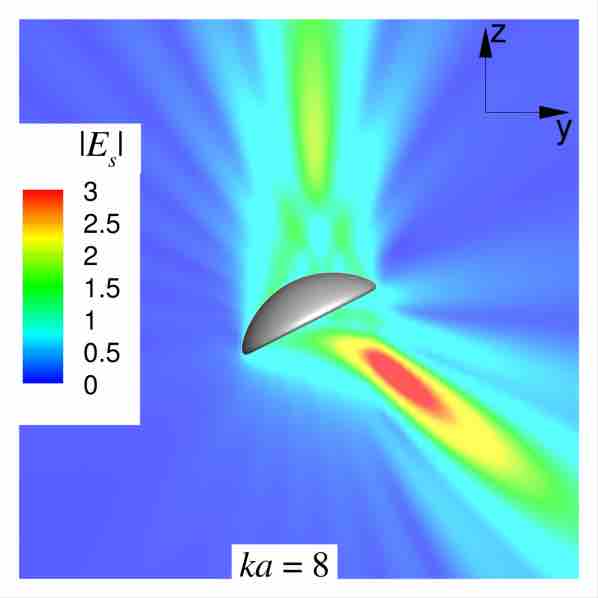}
  \end{tabular} 
  \caption{Scattered electric field amplitudes in the $yz$-plane of an incident plane wave given by Eq.~\ref{eq:E_inc_y} around a perfect electrical conducting sphere or bowl (see text) at the indicated values of the dimensionless wave number, $ka$. The axis of symmetry of the bowl is inclined at an angle $0.15\pi$ radians relative to the direction of propagation, along the positive $z$-axis.}
\label{fig:sphere-bowl-ka}
\end{figure}

\section{Results}

A demonstration of our field only approach for small and large dimensionless wave number, $ka$ where $a$ is a characteristic length scale of the problem, is now given. 
To this end we consider variations of the magnitude of the scattered field around:

a) a perfect electrical conductor (PEC) sphere of radius, $a$,

b) an axisymmetric bowl shaped PEC object obtained by rotating the parametric closed curve in the body coordinates, see Fig.~\ref{fig:SI_bowl-curve}
\begin{equation}
(X, Z) = a~( 2 \sin \theta,~\beta \sin^2 \theta + \gamma [\cos \theta - 1]), \quad 0 \leq \theta < 2 \pi,
\label{eq:bowl_eqn}
\end{equation}
about the $Z$-axis. 
The bowl rim has radius, $2a$ and the inner concave bottom of the bowl can be fitted to a parabola $Z = X^2/(4f)$ where the focal length, $f = R/2$ is related to the radius of curvature, $R = 4a/(2\beta-\gamma)$ at the inner center of the bowl. 
Here we choose $\beta = 0.6$ and $\gamma = 0.5$ that gives $f = 2.86 a$ and $R = 5.71 a$ as shown in Fig.\ref{fig:SI_bowl-curve}. 

c) an axisymmetric thin nano-rod with length, $L$ and maximum cross-section diameter, $b$ whose surface is defined by rotating an analytic curve about the long axis~\cite{Chwang1974}. The aspect ratio of the rod is $L/b = 10$. The cases of the nano-rod being a PEC and a dielectric are studied.

In Fig.~\ref{fig:sphere-bowl-ka} we show the scattered field amplitudes by perfect electrical conductors (PEC): a sphere and a bowl, excited by an incident plane wave that propagates in the positive $z$-direction and with $\boldsymbol{E}$ polarized in the $y$-direction with wave number, $k$.
The results shown in the $yz$-plane demonstrate the transition from the electrostatic limit at small wave numbers, $ka$, to approaching geometric or physical optics as $ka$ increases. 
Note also the high induced local field strengths associated with the corona effect that is evident around the high curvature rim of the bowl in the long wavelength (small $ka$) limit. 

\begin{figure}[t]     
\centering
  \begin{tabular}{@{}cc@{}}
    \includegraphics[width=.23\textwidth]{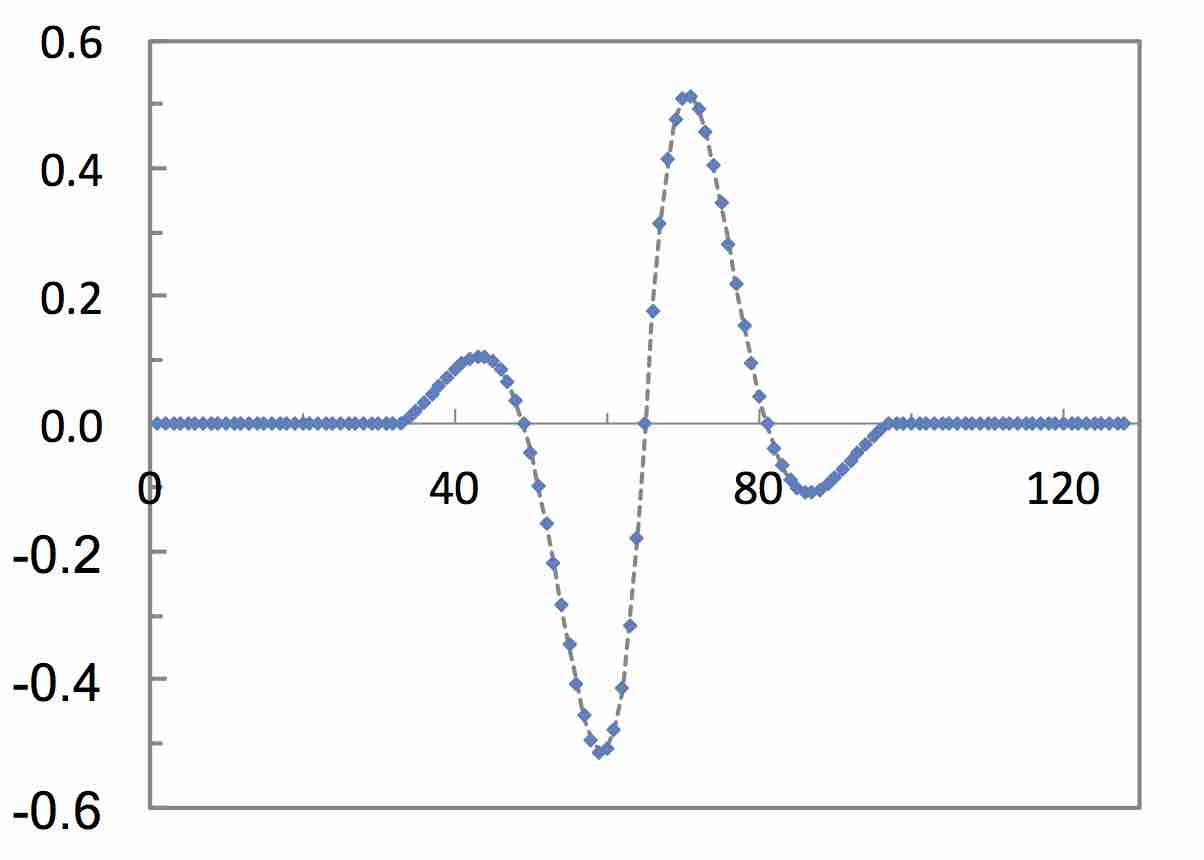}  &
     \includegraphics[width=.23\textwidth]{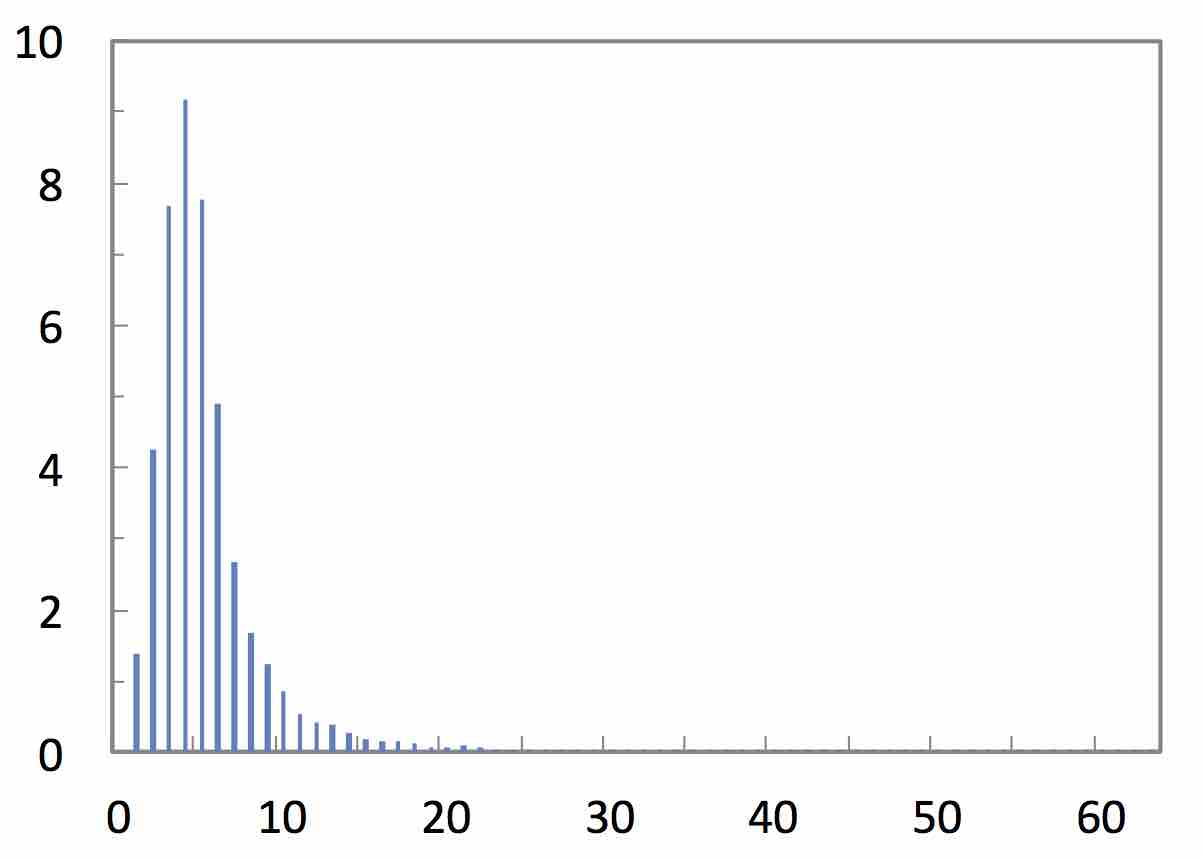} \\
    \includegraphics[width=.23\textwidth]{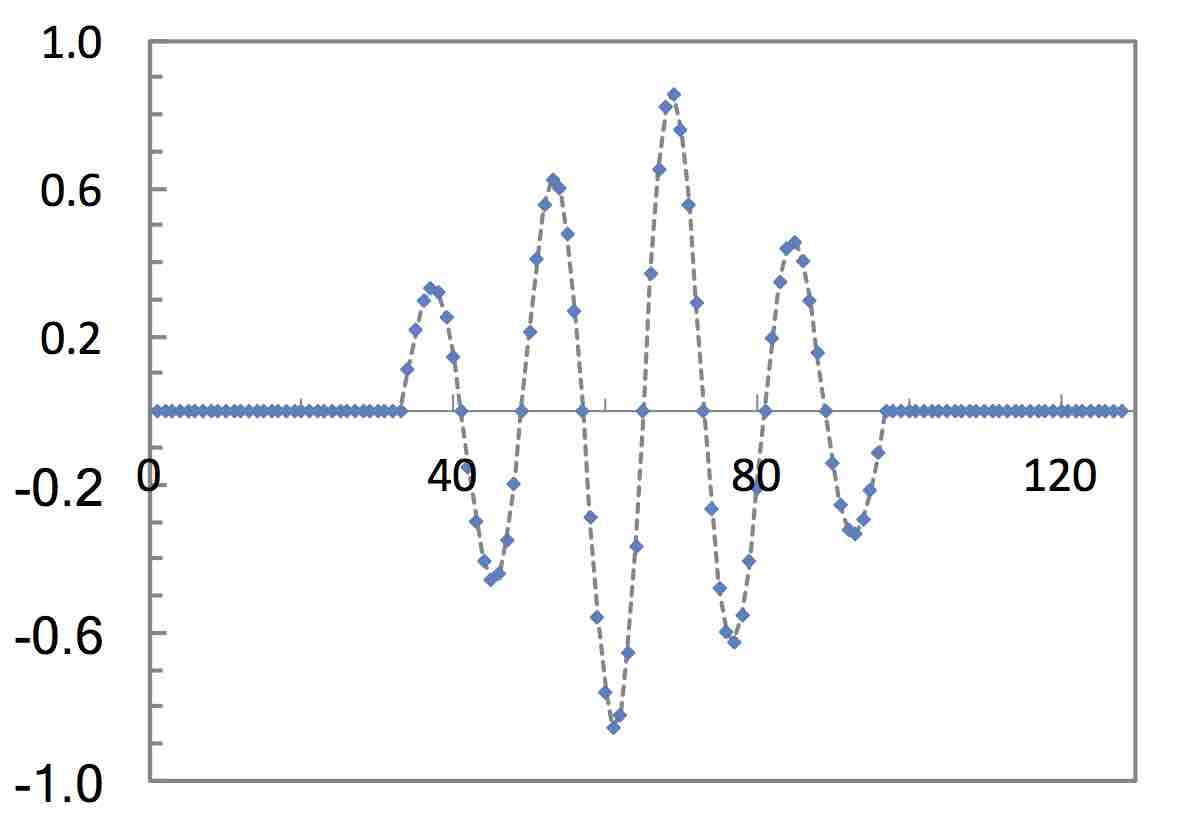} &
     \includegraphics[width=.23\textwidth]{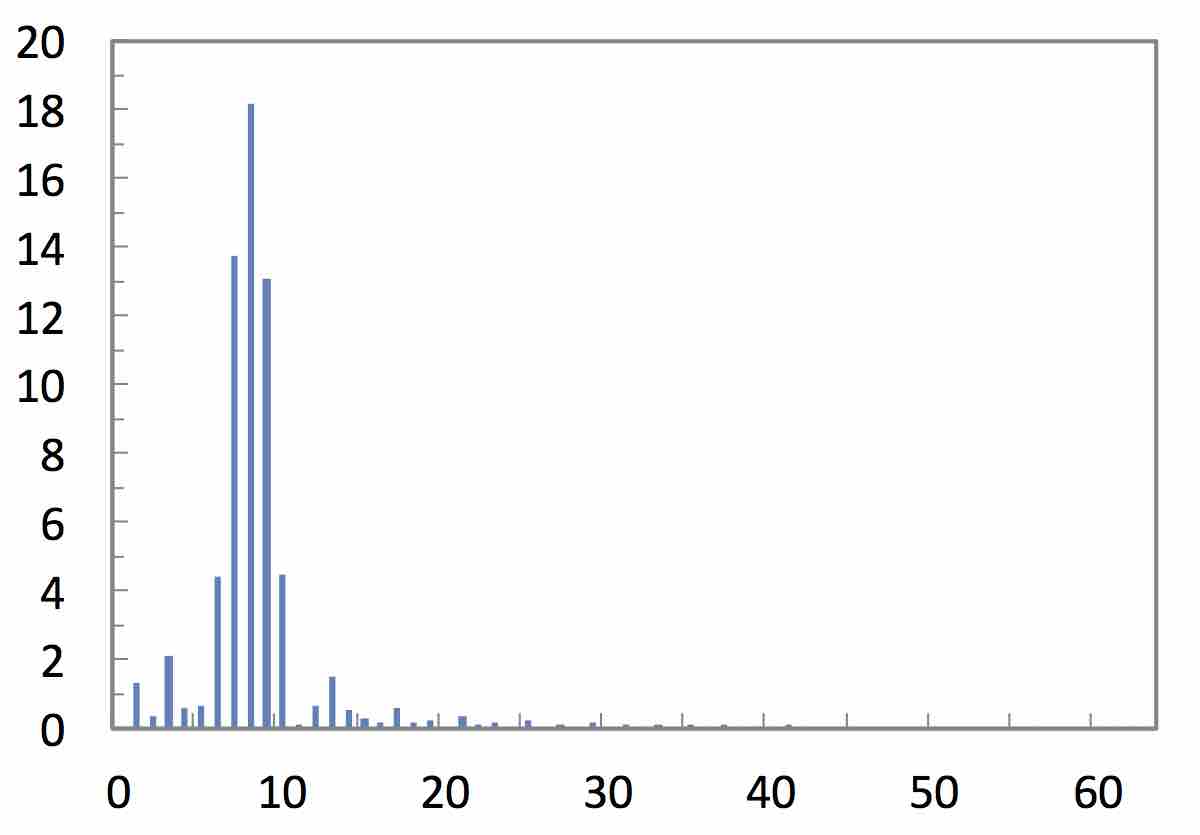} 
  \end{tabular}
\caption{Examples of the incident field pulse $E^{inc}_y$, Eq.~\ref{eq:E_inc_y}, of window width, $w$, and its Fourier amplitudes for pulses that are comprised of (a) 2 cycles, $N_c = 1$, $\alpha = 0.5$ (upper set) and (b) 4 cycles, $N_c = 2$, $\alpha = 0.1$ (lower set) with 128 sampling points and the 64 values of its discrete Fourier transform.}
\label{fig:pulse_FFT}
\end{figure} 

To demonstrate time dependent effects, we consider the scattering of an incident plane wave pulse: $\boldsymbol{E}^{inc} = (0, E^{inc}_y, 0)$ that is polarized in the $y$-direction and propagates in the positive $z$-direction with the following form in a window of width, $w$ 
\begin{eqnarray}
E^{inc}_y = \left\lbrace
  \begin{array} {cccc}
 \qquad 0,  \qquad   \qquad \qquad \qquad -4 \pi N_c \leq \tau  < -2 \pi N_c   \\
\sin(\tau) \exp (-\alpha |\tau|), \qquad    -2 \pi N_c \leq \tau  \leq 2 \pi N_c  \\
\qquad 0,  \qquad  \qquad \qquad \qquad 2 \pi N_c < \tau  \leq 4 \pi N_c     \\
 \end{array} \right.
\label{eq:E_inc_y}
\end{eqnarray}
where $\tau \equiv k_0 (z-vt)$. 
Examples of such pulse functions that comprise of $2N_c$ oscillatory cycles and their Fourier components calculated using a 128 point discrete Fourier transform are given in Fig.~\ref{fig:pulse_FFT} for different values of $N_c$ and $\alpha$~\cite{Press1992}. 
Since the Fourier amplitudes are small at high wave numbers, using say the first 20 Fourier components is generally sufficient to give an accurate representation of the pulse.

For each of these Fourier components of the incident pulse we solve for $\boldsymbol{E}$ and $(\boldsymbol{x} \cdot \boldsymbol{E})$ at the corresponding value of $k$ and then the time domain behavior can be found by inverse discrete fast Fourier transform~\cite{Press1992}.  
As in earlier work on the scattering of acoustic pulses~\cite{Klaseboer2017b}, we set the total window width of the incident pulse to be $w = 20.1a$, of which the central oscillatory portion has width $w/2$. 
So for a pulse with $2N_c$ oscillatory cycles over the width $w/2$, the wave number, $k_0$ in Eq.~\ref{eq:E_inc_y} is determined by the relation $(2N_c)\lambda_0 \equiv (2N_c) (2\pi/k_0) = w/2$, giving $k_0 a = 8\pi N_c/20.1$. The results are presented in such a way, that the pulse just reaches the bottom of the scatterer at the first frame. To account for this, each frequency component of the incoming wave must be multiplied with a phase factor $e^{i\beta}$ with $\beta=-k_ma[w/(4a)-1]$ and $k_m$ the $m^\text{th}$ wave number of the Fourier spectrum. 

\begin{figure}[t]   
\centering{}
\begin{tabular}{@{}cc@{}}
\includegraphics[width=.23\textwidth]{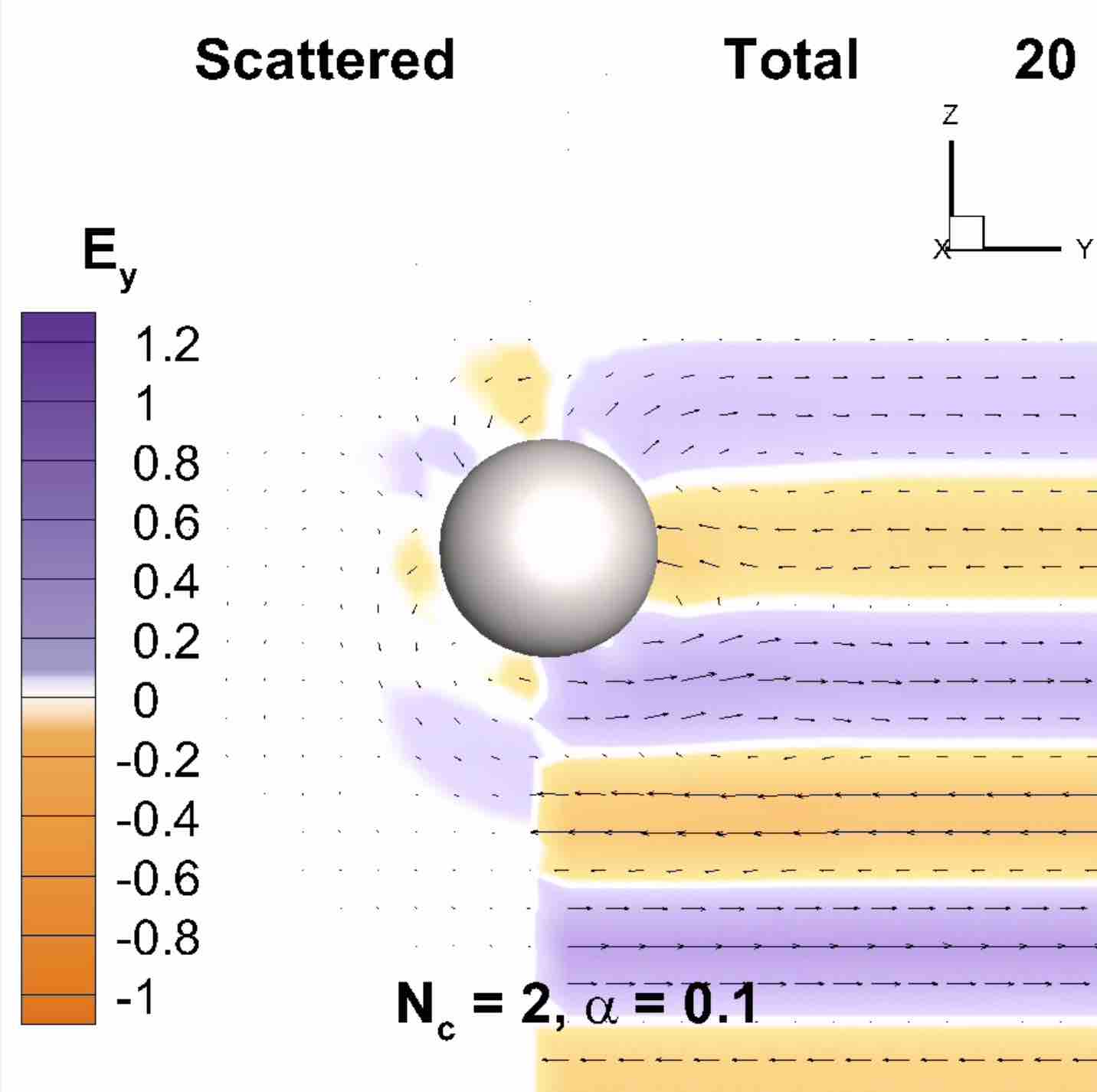} & 
\includegraphics[width=.23\textwidth]{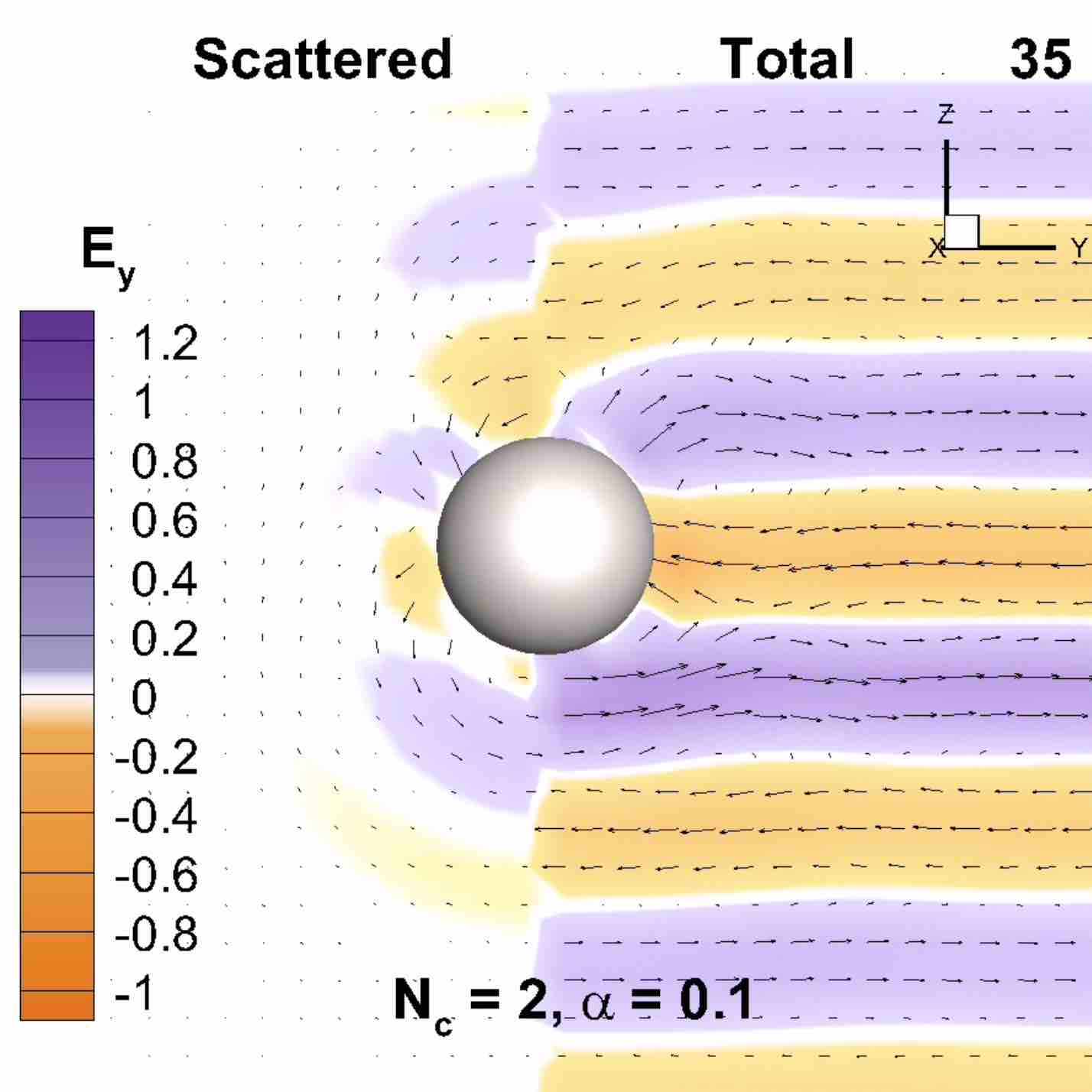} \\ 
\includegraphics[width=.23\textwidth]{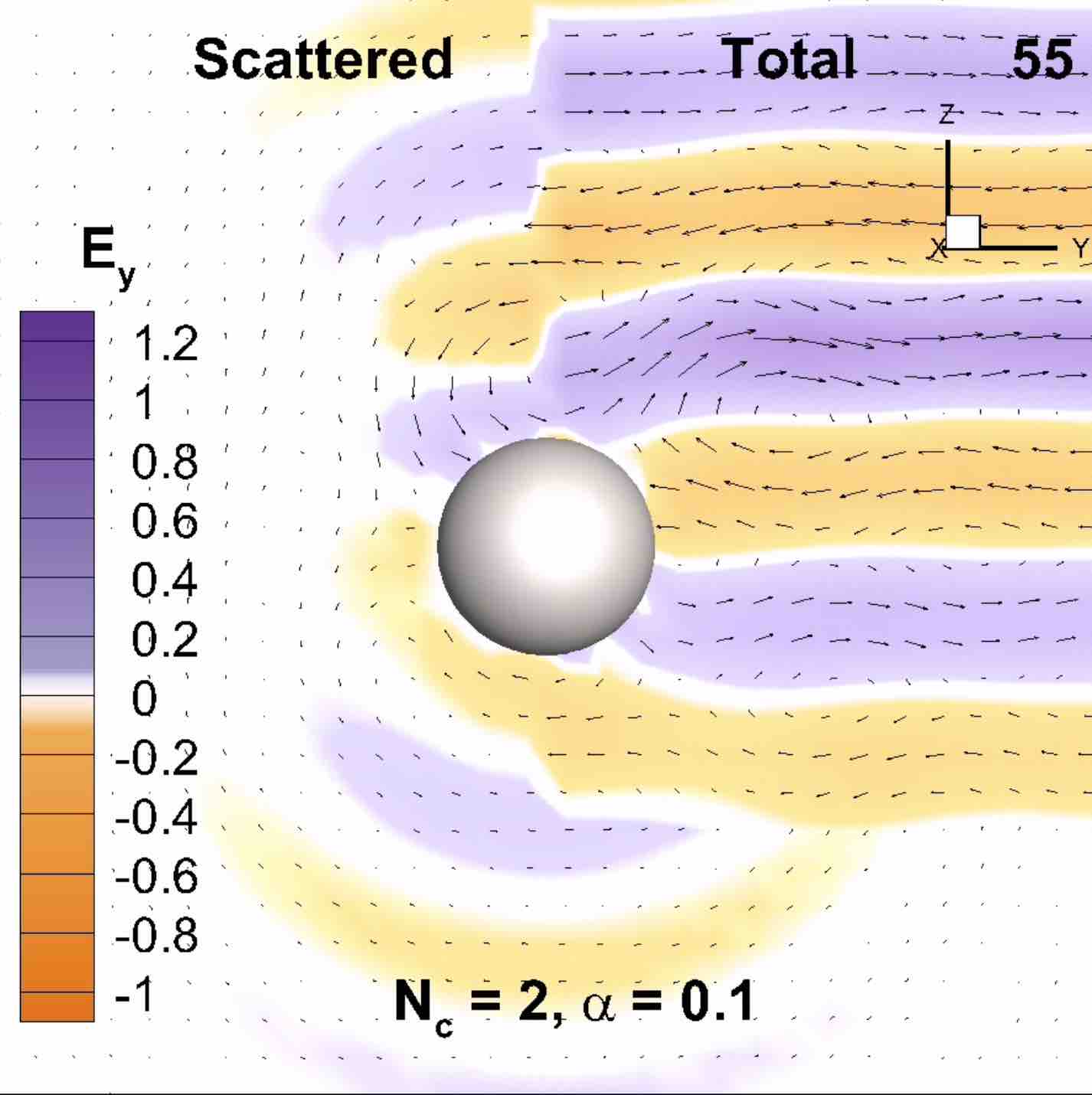} & 
\includegraphics[width=.23\textwidth]{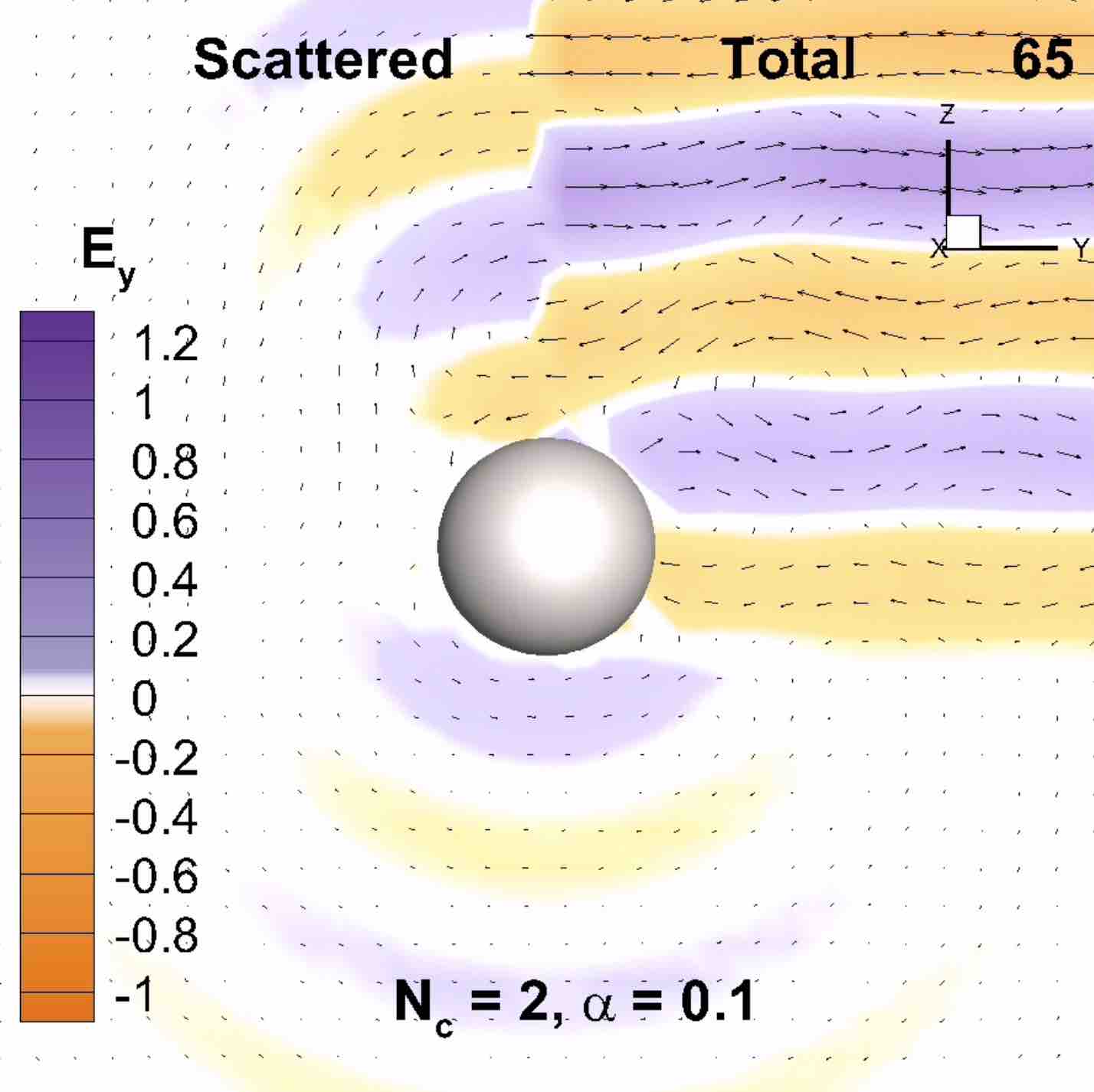}
\end{tabular}
  \caption{Total (right) and scattered (left) electric field amplitudes and field vectors in the $yz$-plane due to scattering of the plane wave pulse, Eq.~\ref{eq:E_inc_y} with $k_0 a =16\pi/20.1$, by a perfect electrical conducting sphere at selected time steps. See Visualization 1 in the Supplementary Material for full animation.}
\label{fig:sphere-pulse}
\end{figure}

In Fig.~\ref{fig:sphere-pulse} we show both the total and scattered $E_y$ electric field amplitudes as well as the electric field vectors excited by the plane wave pulse given by Eq.~\ref{eq:E_inc_y}. Results in the $yz$-plane around a PEC sphere are shown at selected time steps out of a total of 128 steps (1002 nodes and 500 quadratic elements were used). The field strengths are indicated by the color scale and their strengths and directions by arrows. These results have been validated against available analytic solutions~\cite{Liou1977}.

\begin{figure}[t]       
\centering
\begin{tabular}{@{}cc@{}}
\includegraphics[width=.23\textwidth]{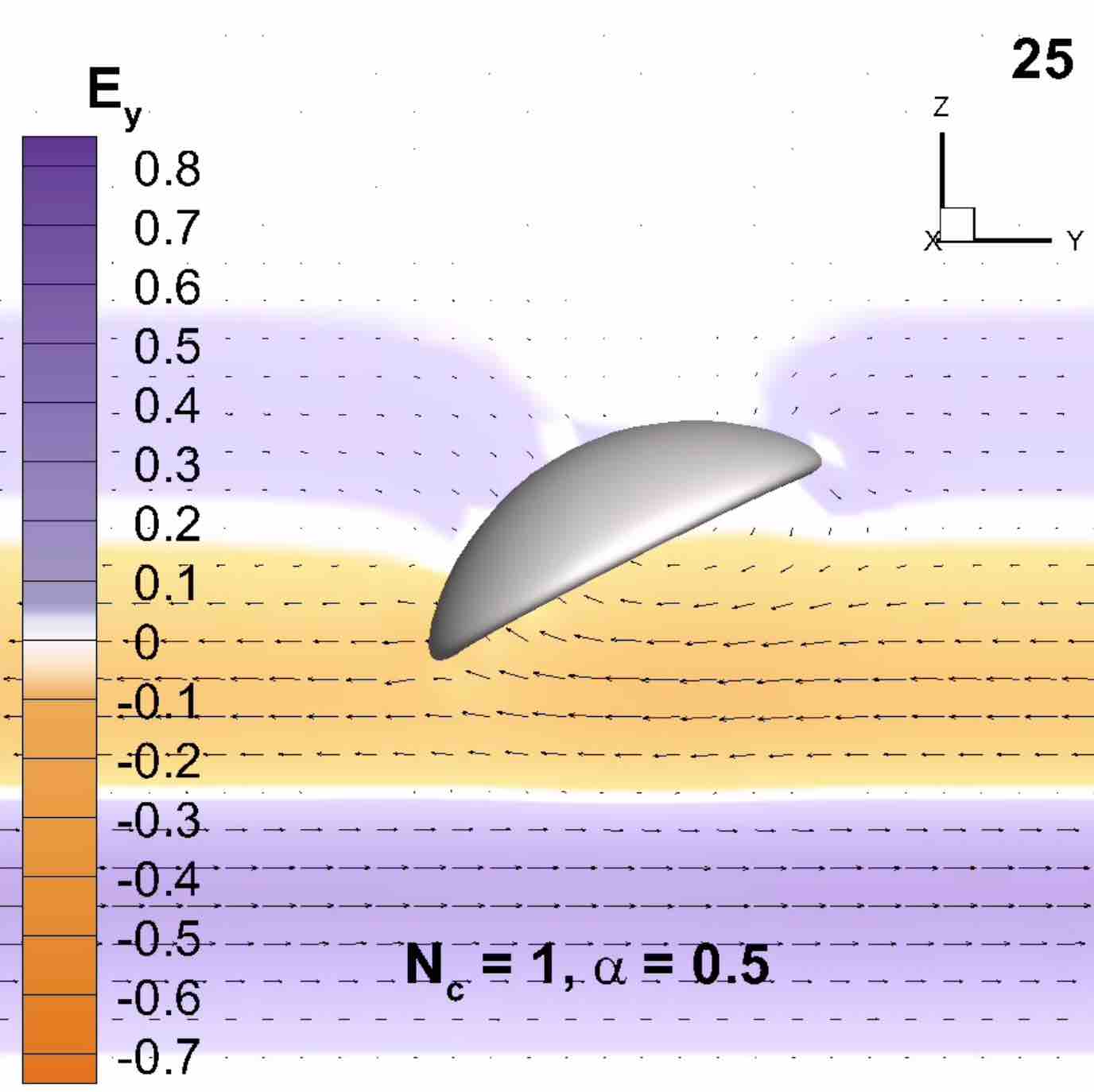} &
\includegraphics[width=.23\textwidth]{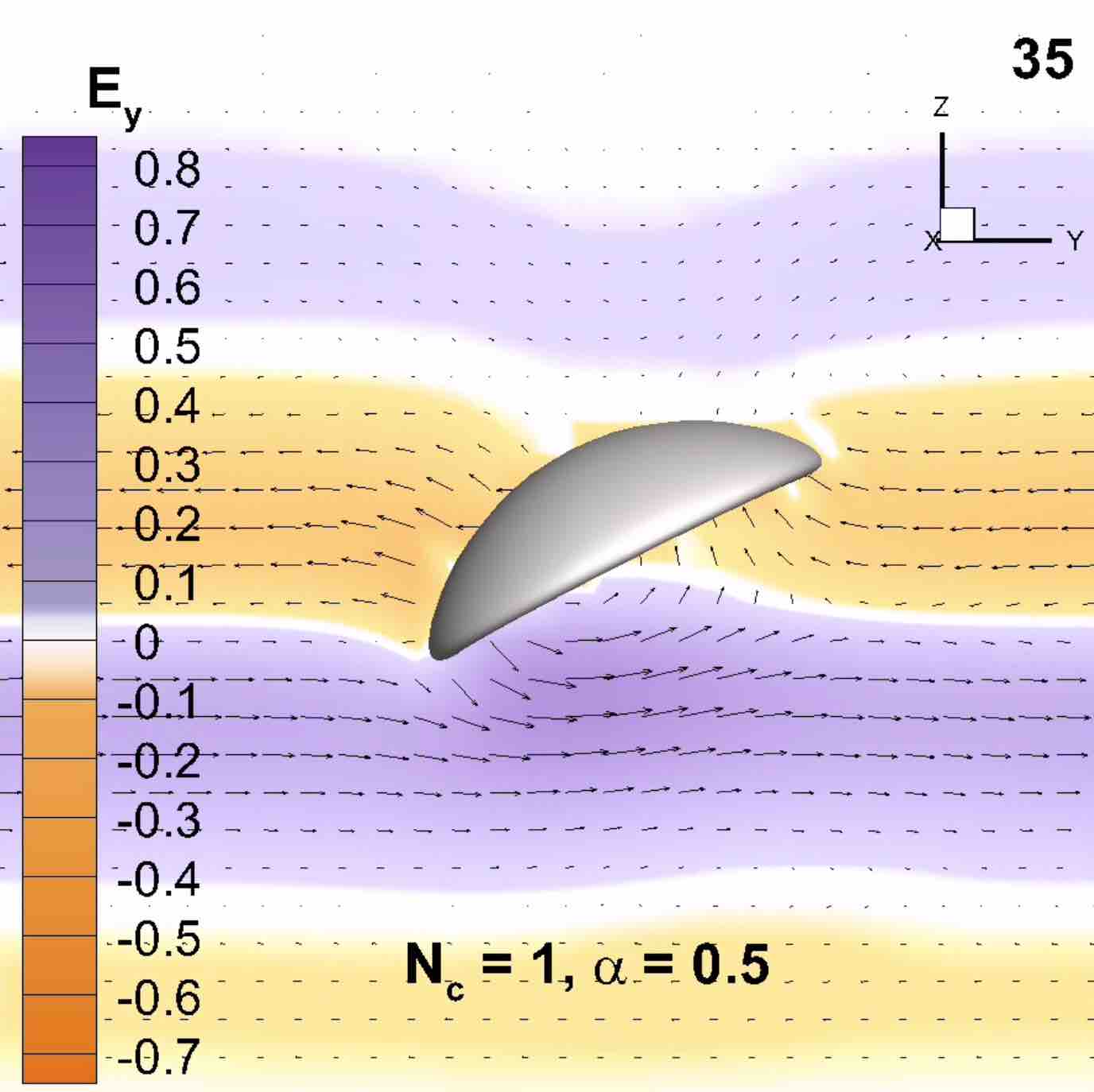}  \\ 
\includegraphics[width=.23\textwidth]{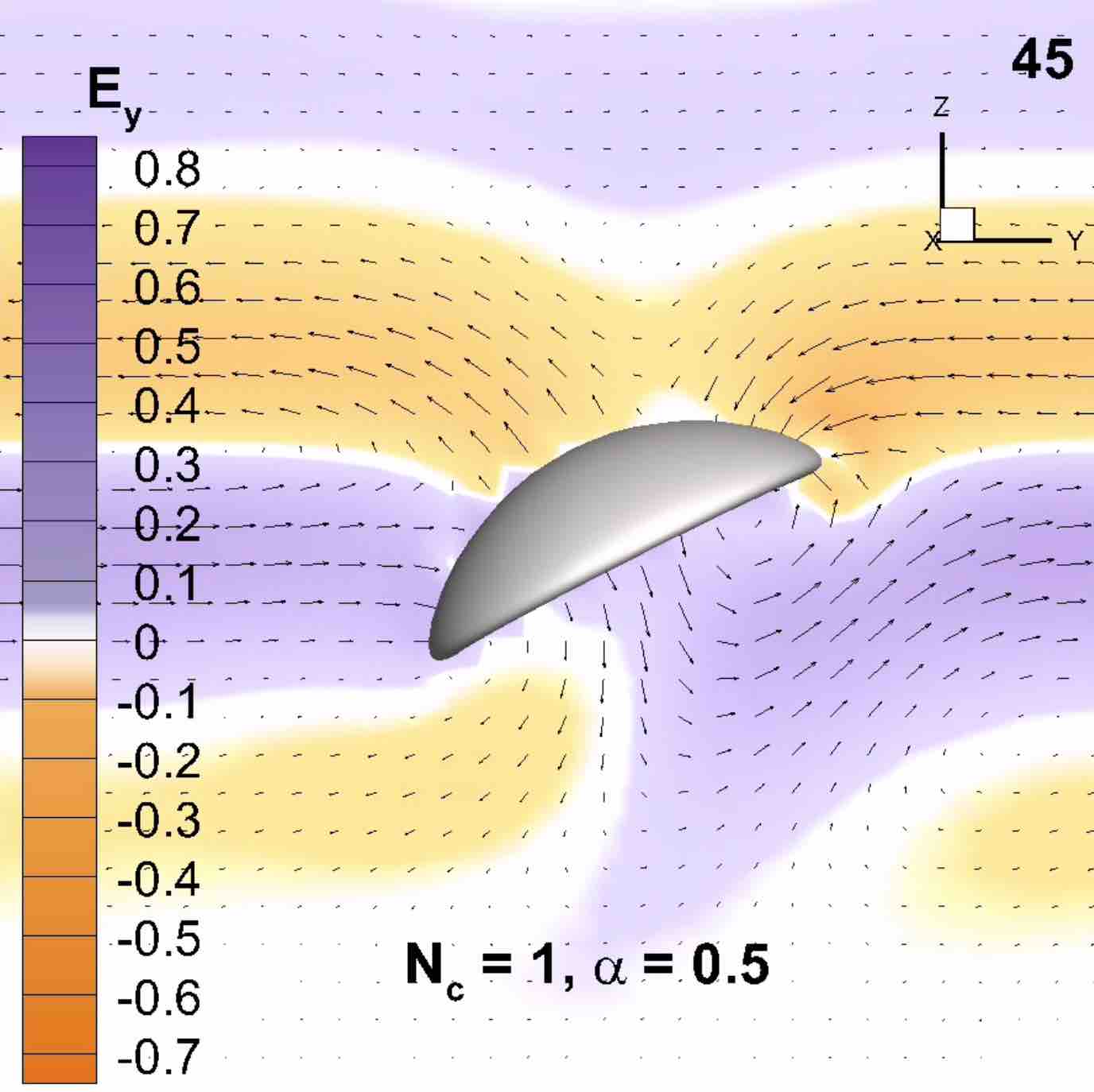} &
\includegraphics[width=.23\textwidth]{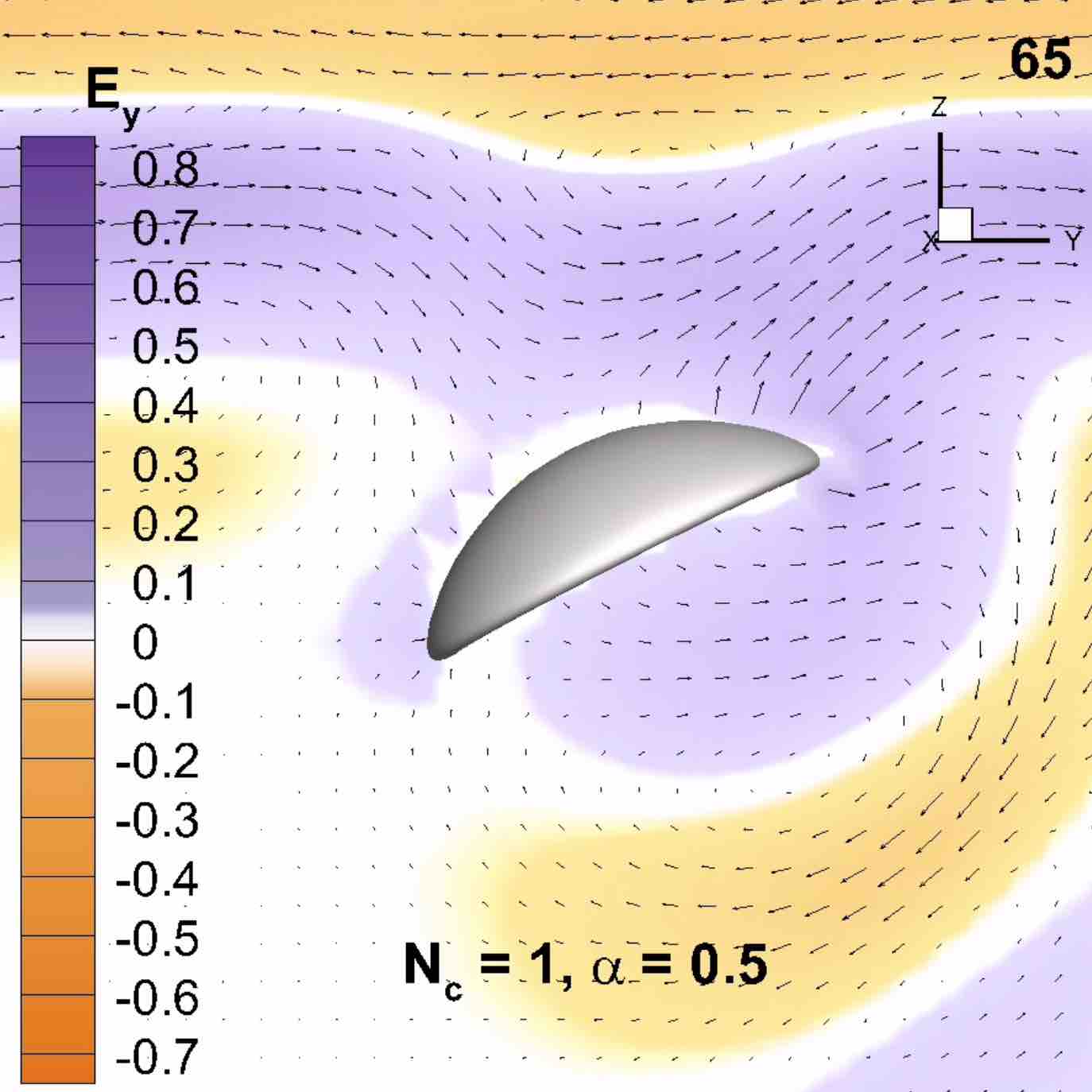}
\end{tabular}
  \caption{Snapshots of the total $\boldsymbol{E}$ field amplitude and field vector in the $yz$-plane due to scattering of a plane wave pulse given by Eq.~\ref{eq:E_inc_y} with $k_0 a = 8\pi/20.1$ by a perfect electrical conducting bowl of rim radius $2a$ at the indicated time steps. The bowl is inclined at an angle $0.15\pi$ radians relative to the $z$-axis. See Visualization 2 in the Supplementary Material for full animation.}
\label{fig:Bowl-pulse-SinExp-Nc1-A05}
\end{figure}

\begin{figure}[t]       
\centering
\includegraphics[width=.23\textwidth]{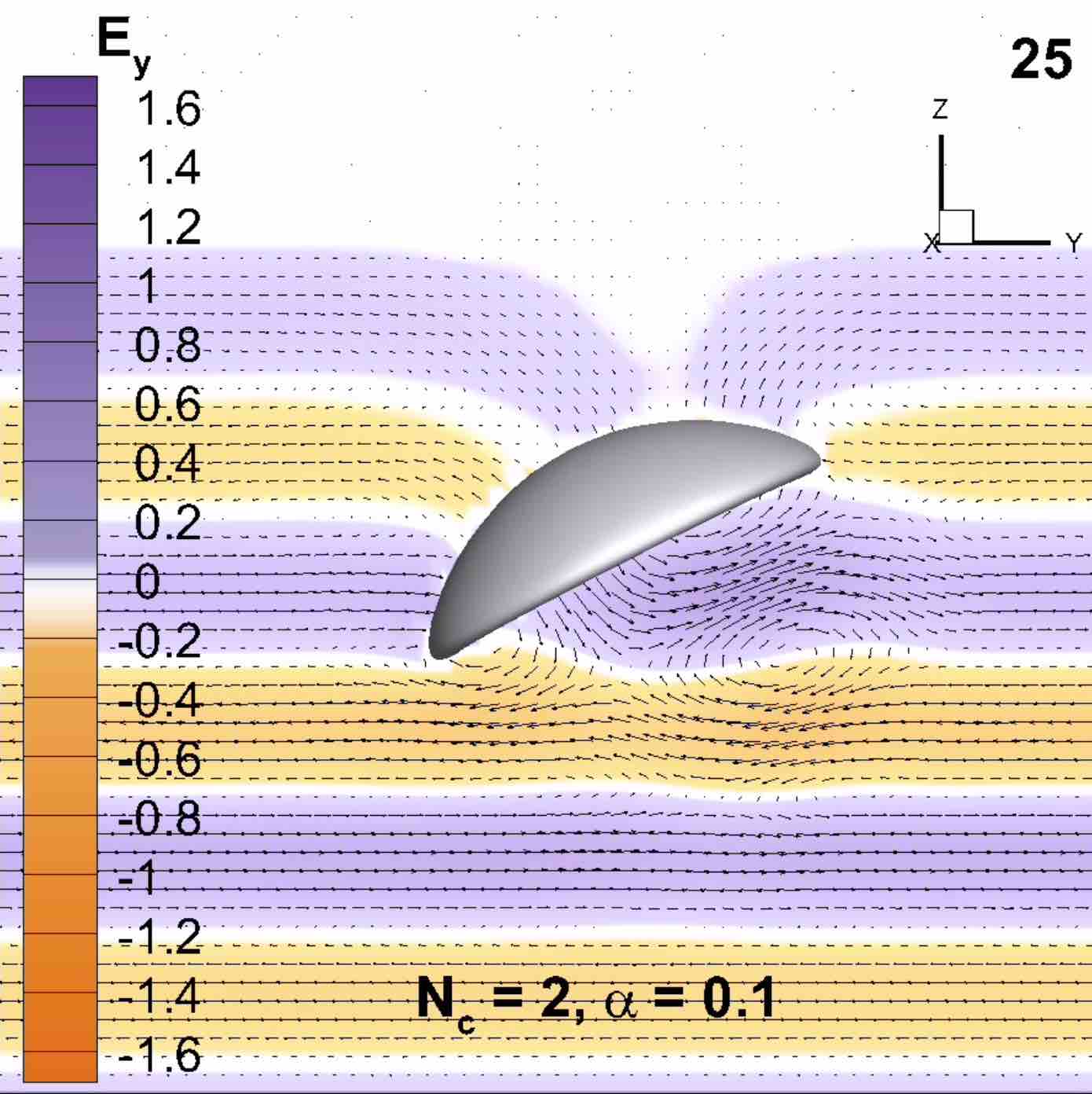} \; \includegraphics[width=.23\textwidth]{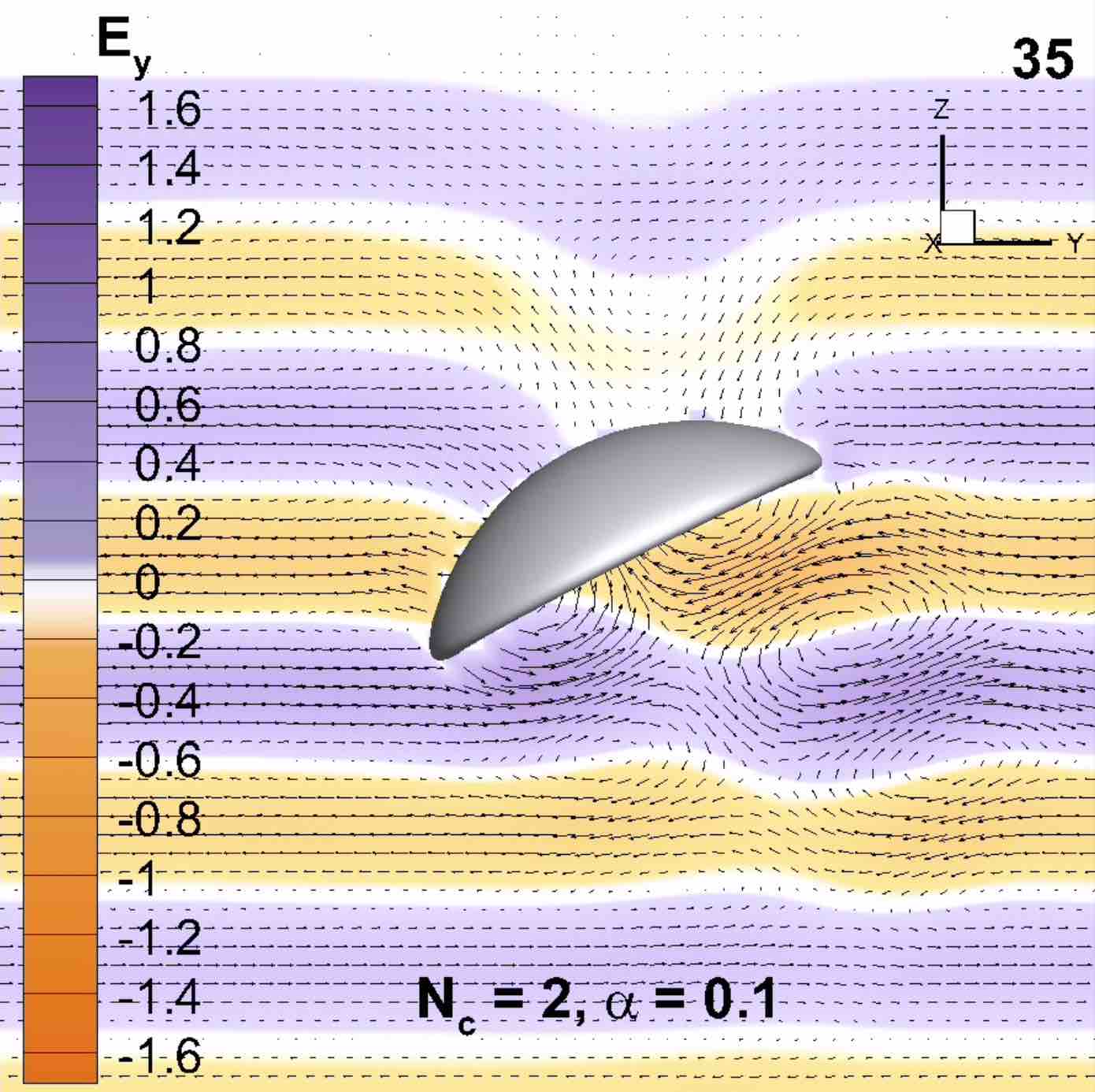} \\
\medskip
\includegraphics[width=.23\textwidth]{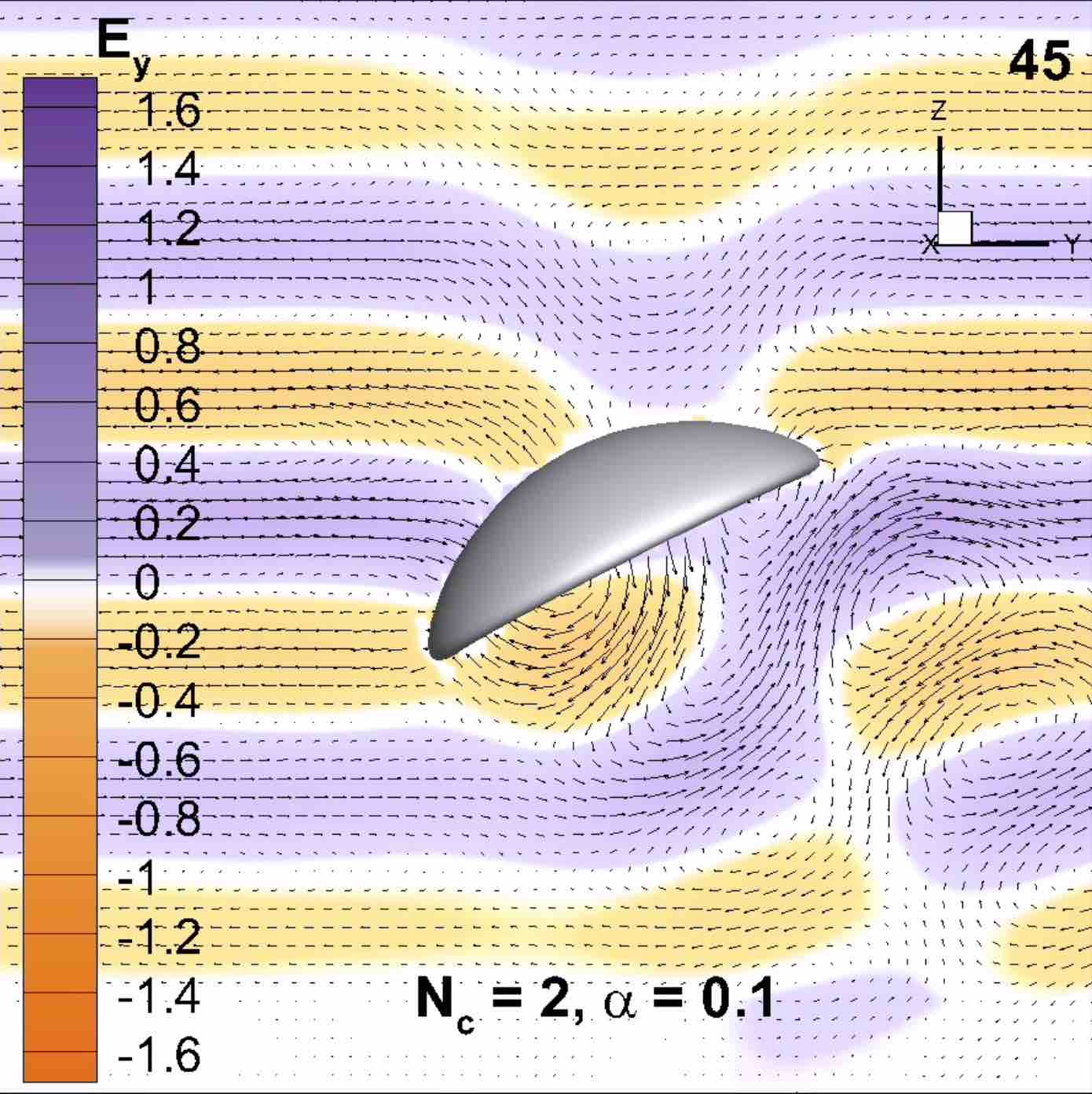}\;
\includegraphics[width=.23\textwidth]{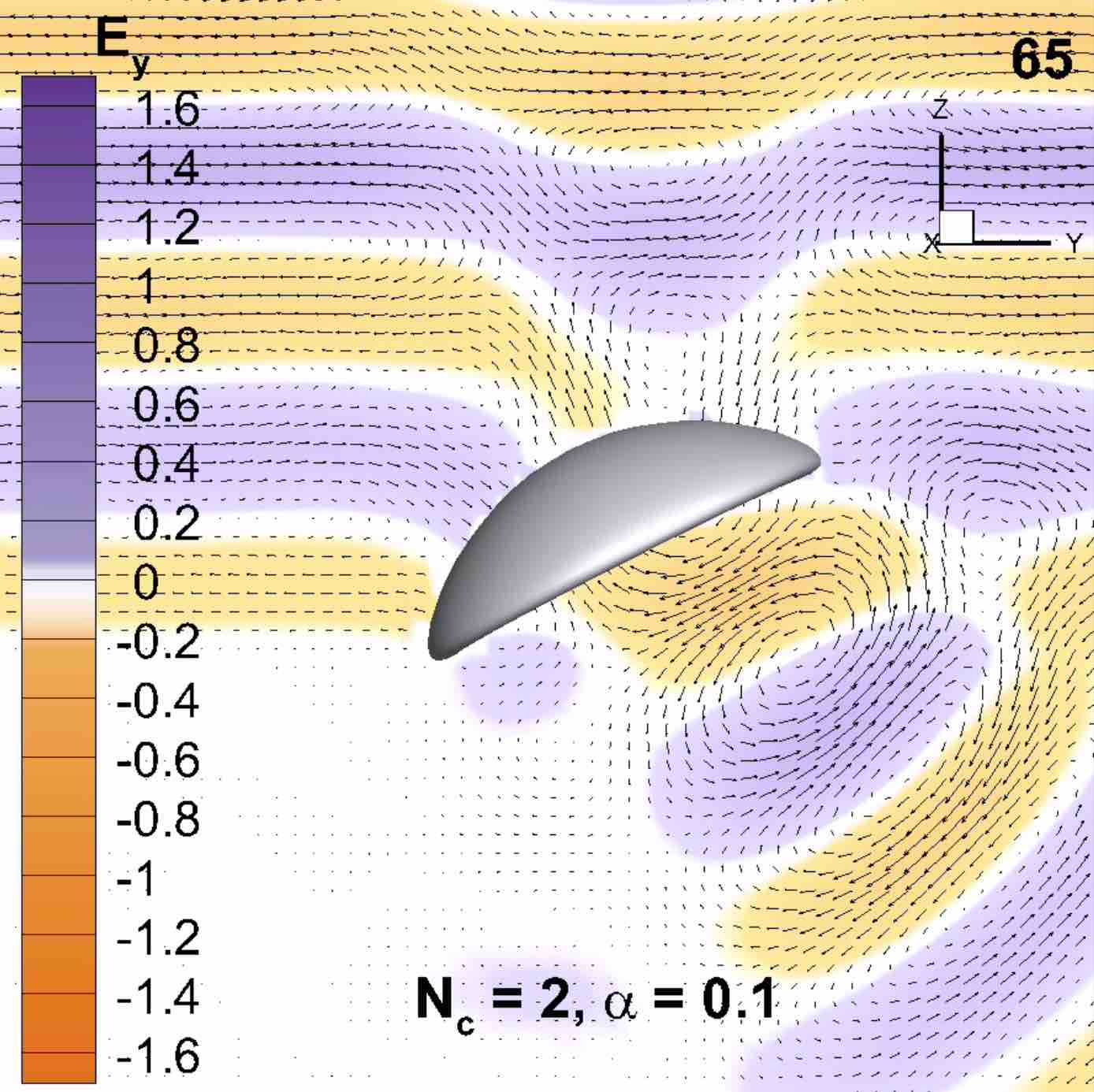} \\
\medskip
\includegraphics[width=.23\textwidth]{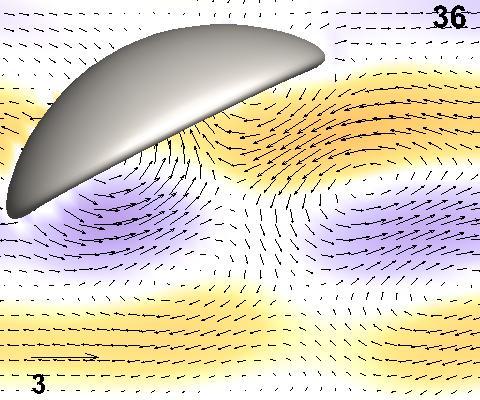}\;
\includegraphics[width=.23\textwidth]{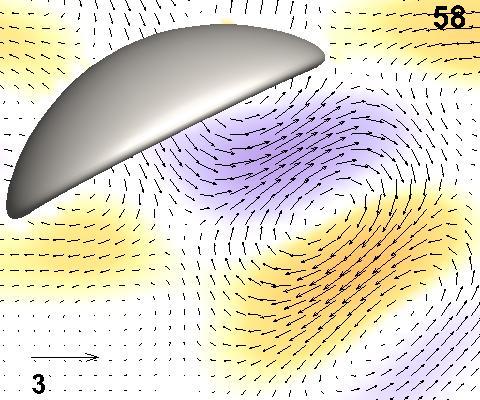}
  \caption{Snapshots of the total $\boldsymbol{E}$ field amplitude and field vector in the $yz$-plane due to scattering of a plane wave pulse by a perfect electrical conducting bowl of rim radius $2a$. The incident wave is given by Eq.~\ref{eq:E_inc_y} with $k_0 a = 16\pi/20.1$ at the indicated time steps. The bowl is inclined at an angle $0.15\pi$ radians relative to the $z$-axis. Also shown are a magnified region to show the local structure of the field vectors. See Visualization 3 in the Supplementary Material for full animation.}
\label{fig:Bowl-pulse-SinExp-Nc2-A01}
\end{figure}

In Fig.~\ref{fig:Bowl-pulse-SinExp-Nc1-A05}, we show the magnitudes of the total field excited by a broader incident plane wave pulse given by Eq.~\ref{eq:E_inc_y} with $N_c = 1$, $\alpha = 0.5$ and hence $k_0 a = 8\pi/20.1$ in the neighborhood of a PEC bowl with 1002 nodes and 500 quadratic elements.
Here the high field strengths associated with the corona effect are more evident around the high curvature rim of the bowl although the maximum field amplitude is only about half that in Fig.~\ref{fig:sphere-pulse}.

In Fig.~\ref{fig:Bowl-pulse-SinExp-Nc2-A01} we show the total amplitudes and field vectors of $\boldsymbol{E}$ excited by the same incident plane wave pulse as in Fig.~\ref{fig:sphere-pulse} in the neighborhood of a PEC bowl with rim radius $2a$ as given in Eq.~\ref{eq:bowl_eqn}. 
The electric field strengths are indicated by the color scale and the field strengths and directions by arrows.
The focusing effects due to the curvature of the bowl are clearly evident with regions of high field amplitudes located near the position of the focal point discussed above. 
The focusing effect increases the field amplitude by up to about 50\%. 
Complex structures in the vector $\boldsymbol{E}$ are evident in the enlarged sub-figures near the concave surface of the bowl.

\begin{figure}[t]   
\centering
  \begin{tabular}{@{}cc@{}}
    \includegraphics[width=.23\textwidth]{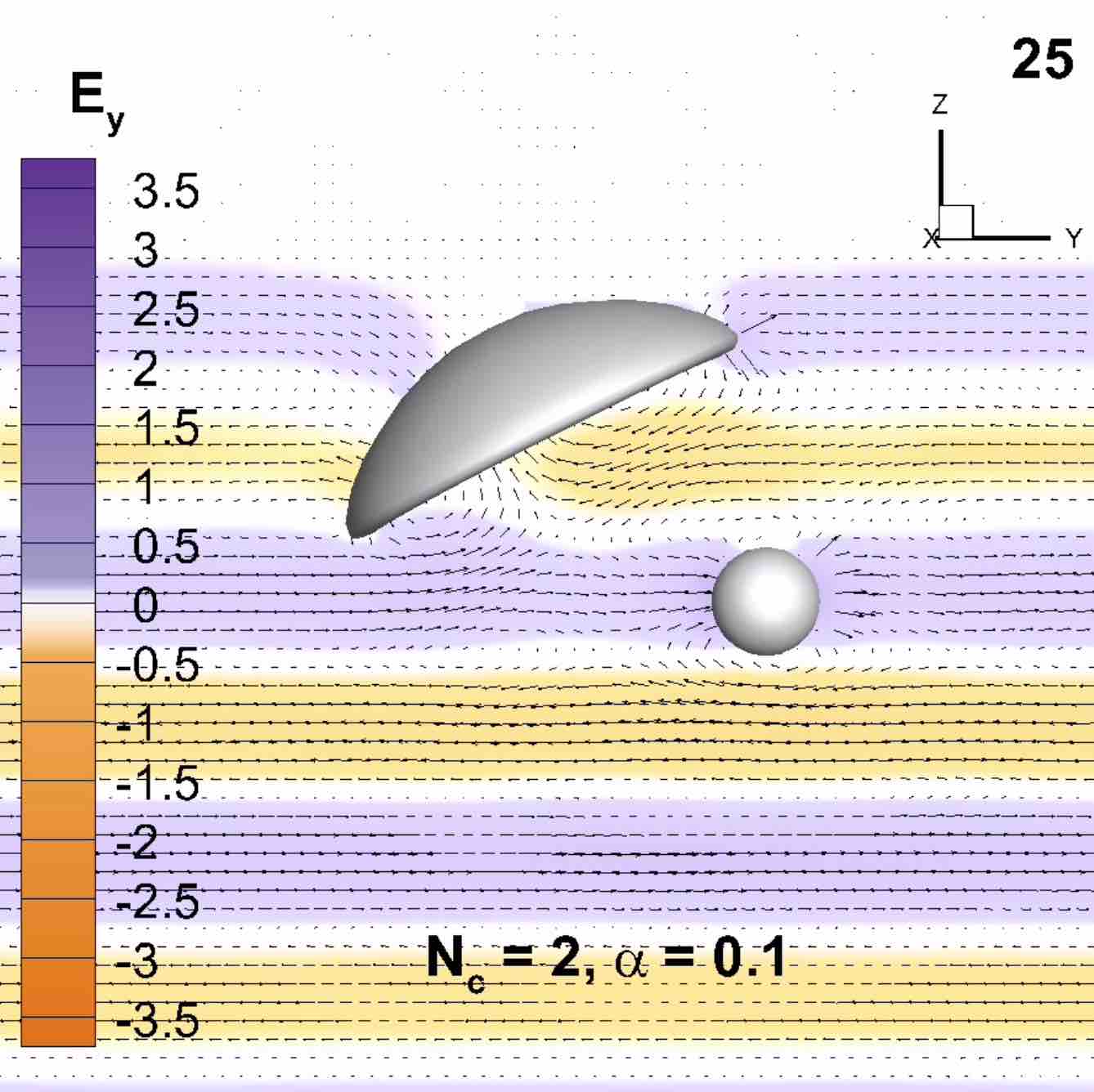} &
    \includegraphics[width=.23\textwidth]{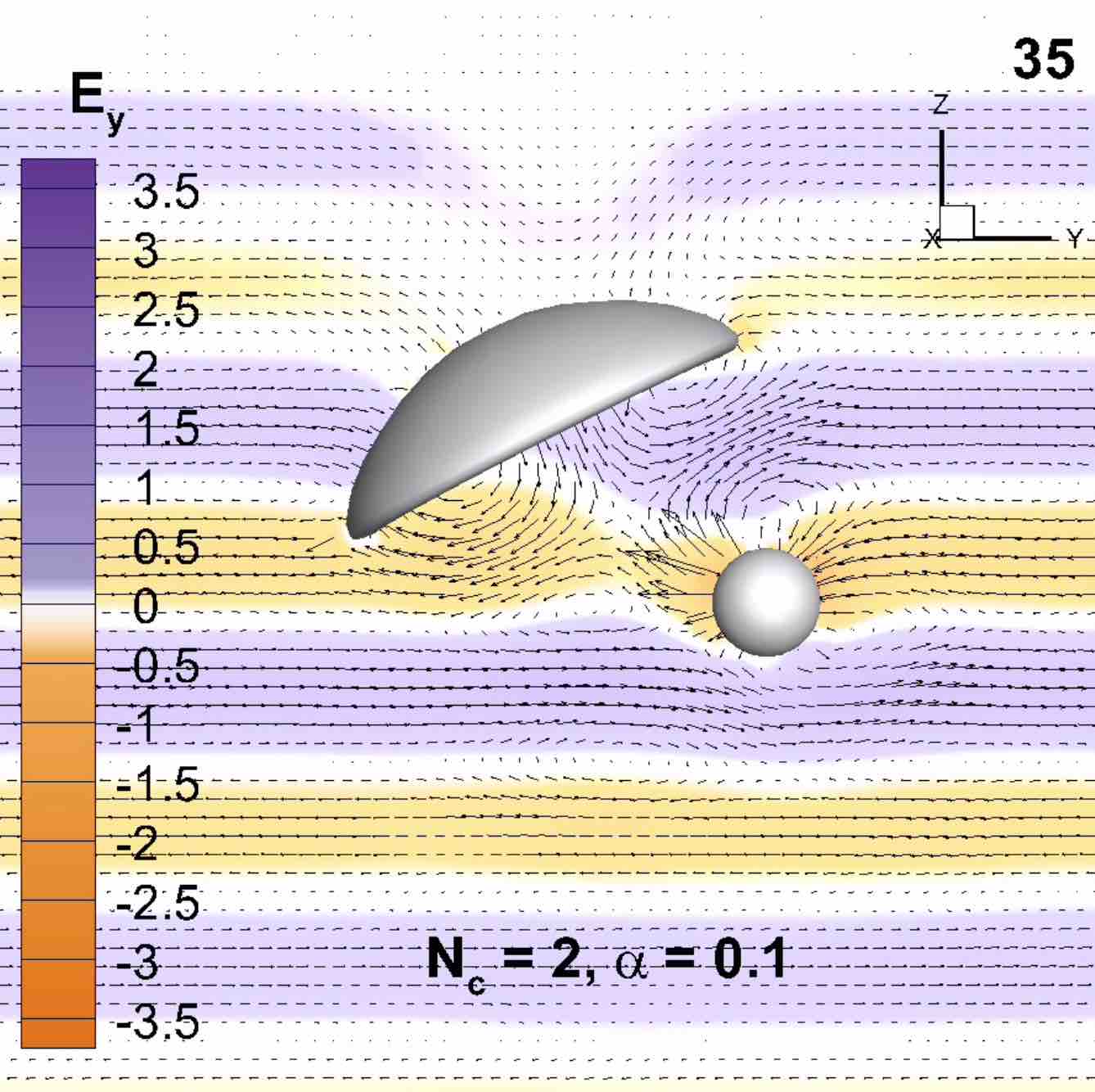} \\
    \includegraphics[width=.23\textwidth]{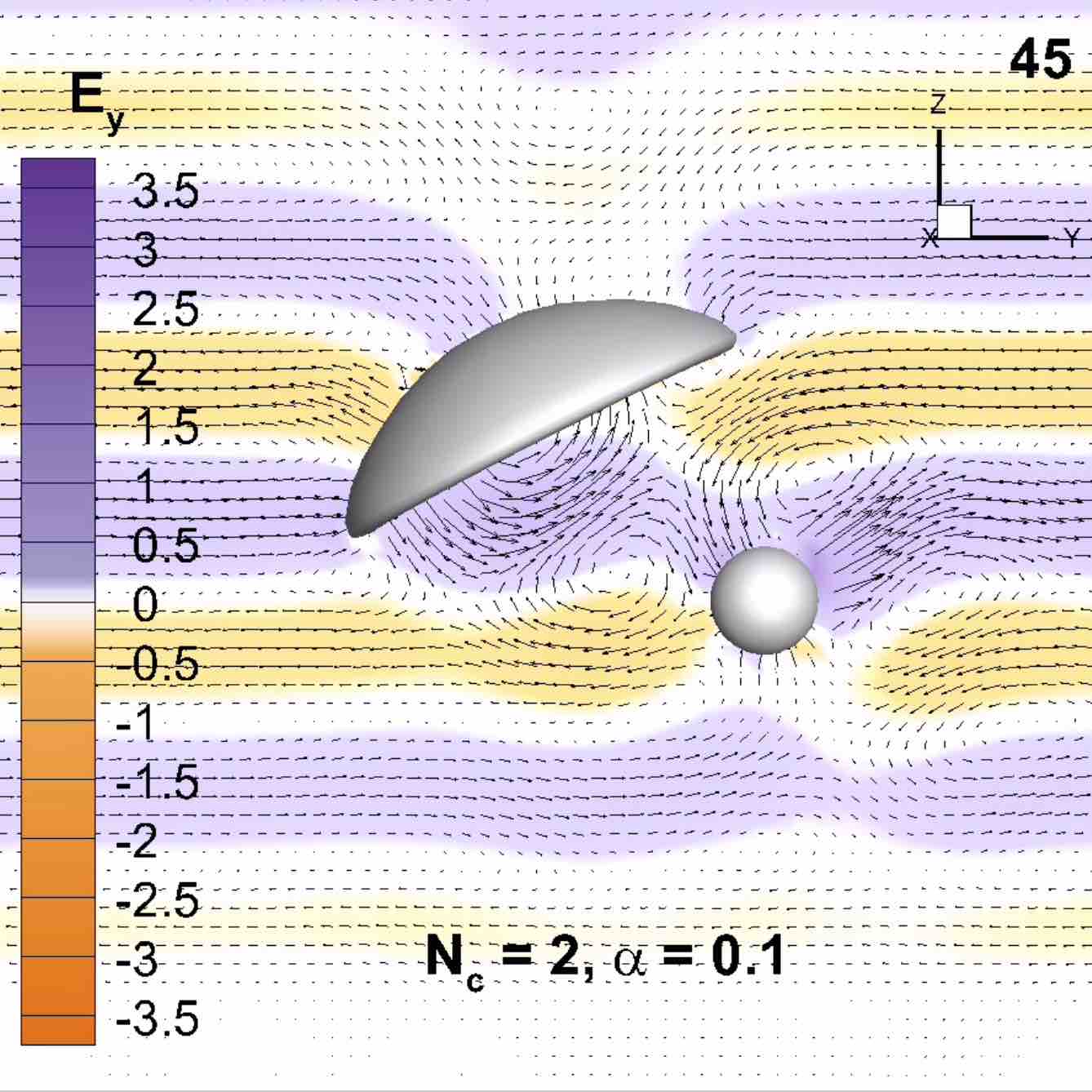} &
    \includegraphics[width=.23\textwidth]{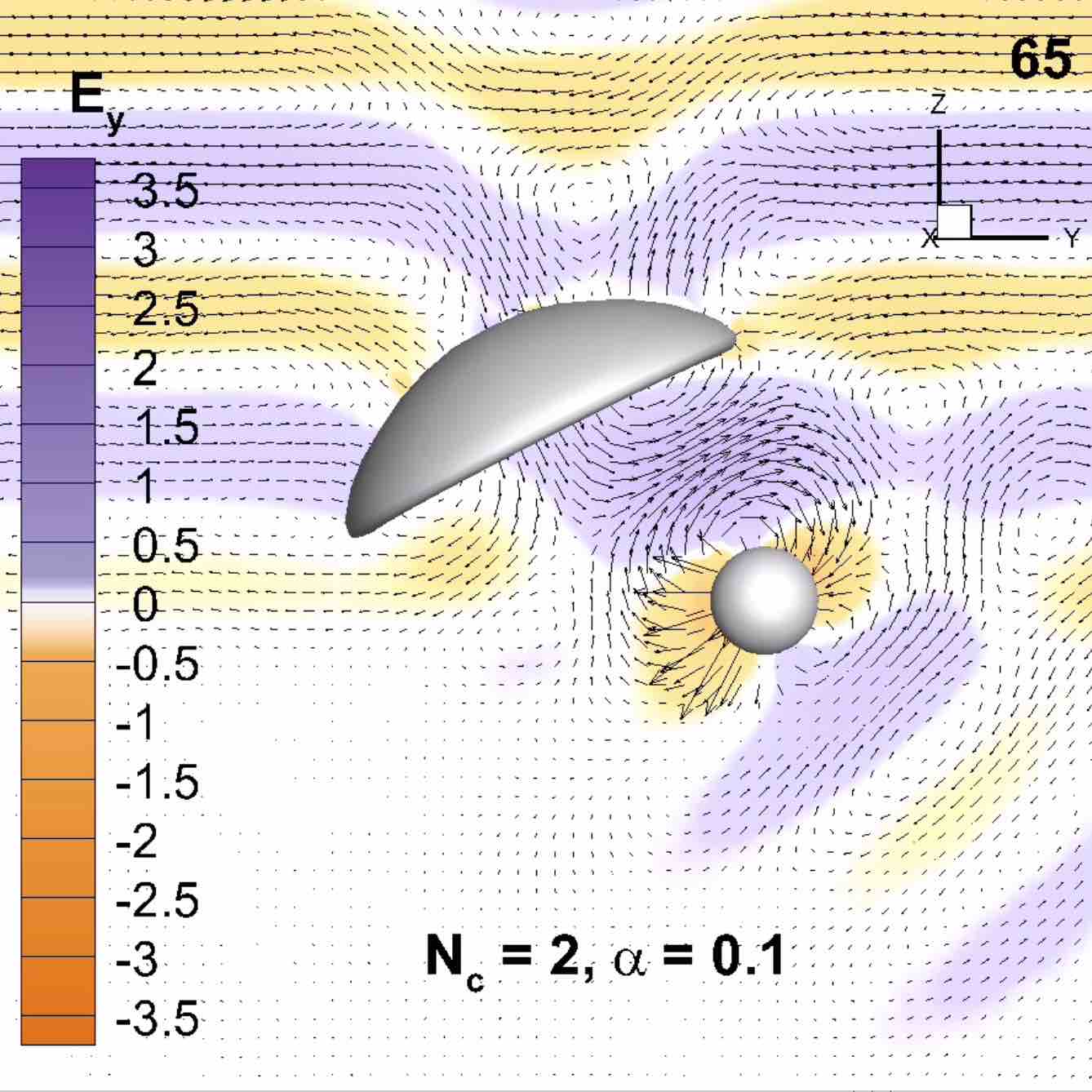} \\
    \includegraphics[width=.23\textwidth]{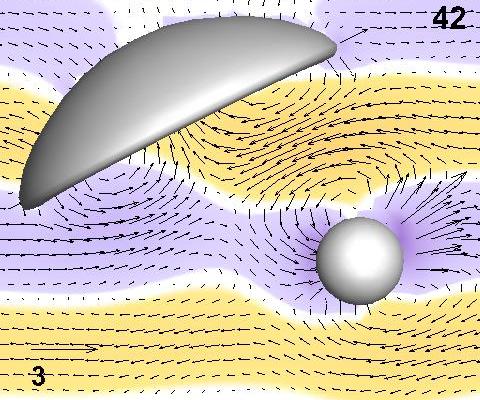} &
    \includegraphics[width=.23\textwidth]{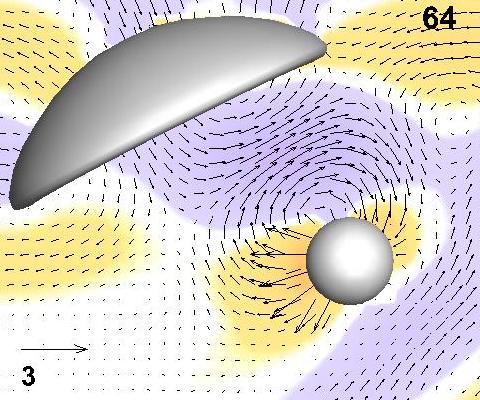}
  \end{tabular}
  \caption{Same as Fig.~\ref{fig:Bowl-pulse-SinExp-Nc2-A01}, but with the addition of a sphere positioned near the focus of the bowl. See Visualization 4 in the Supplementary Material for full animation.}
\label{fig:Bowl-and-sphere}
\end{figure}

To demonstrate the ability of the present field only formulation to handle multiple scattering effects in which spatial configurations of the scatterer can cause significant field enhancements, we consider the scattering by the combined effect of a PEC bowl that has a rim radius, $2a$ with a sphere of diameter $a$, that is located near the focal point of the bowl. The bowl and the sphere each have 362 nodes and 180 quadratic elements.
In Fig.~\ref{fig:Bowl-and-sphere}, we show the space-time variation of the total field amplitudes and directions, for an incident pulse given by Eq.~\ref{eq:E_inc_y} with $N_c = 2$, $\alpha = 0.1$ and $k_0 a = 16\pi/20.1$ -- the same as that in Fig.~\ref{fig:Bowl-pulse-SinExp-Nc2-A01}. Here we see that this configuration of a bowl with a sphere placed near its focus can enhance the local field amplitude by about a factor of 3, twice that achieved by a bowl on its own. The complex field structures in the region between the bowl and the sphere are shown in the enlarged sub-figures.

\begin{figure}[t]   
\centering
  \begin{tabular}{@{}cc@{}}
    \includegraphics[width=.23\textwidth]{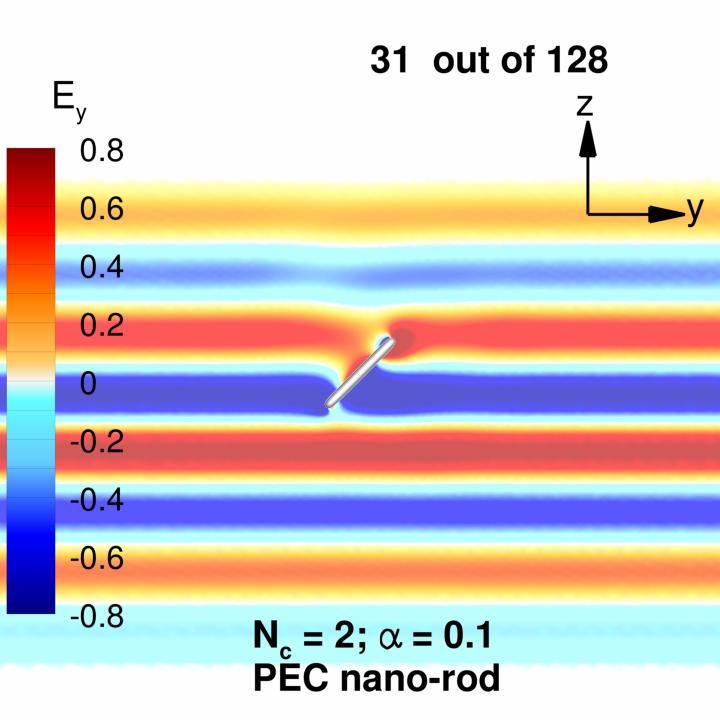} &
    \includegraphics[width=.23\textwidth]{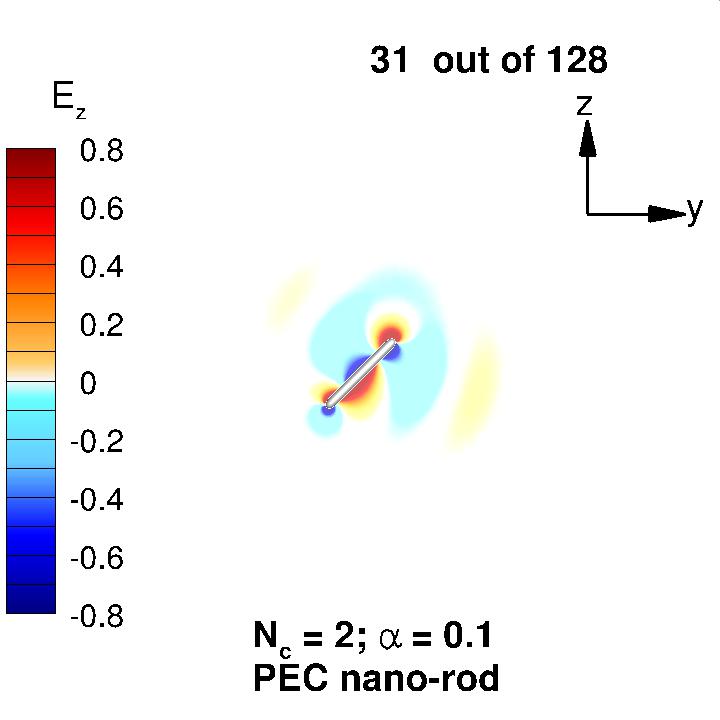} \\
    \includegraphics[width=.23\textwidth]{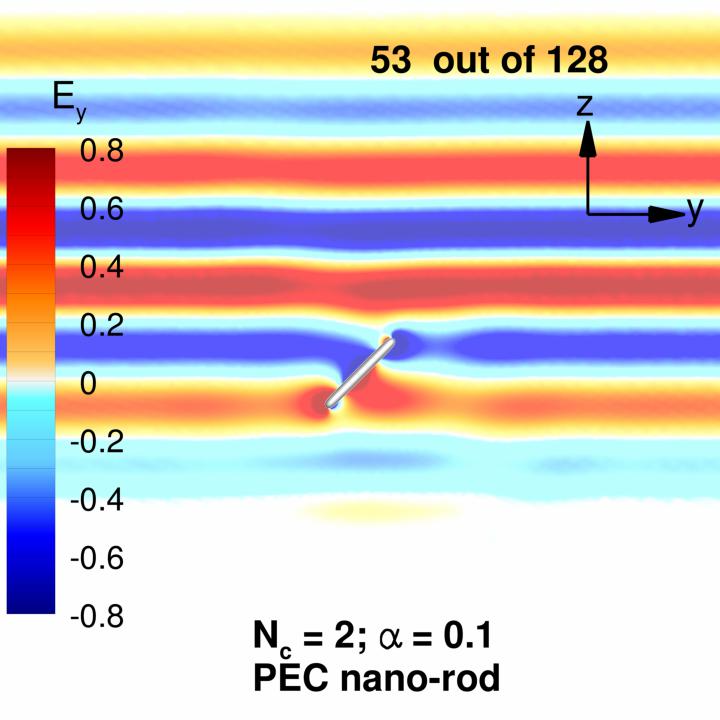} &
    \includegraphics[width=.23\textwidth]{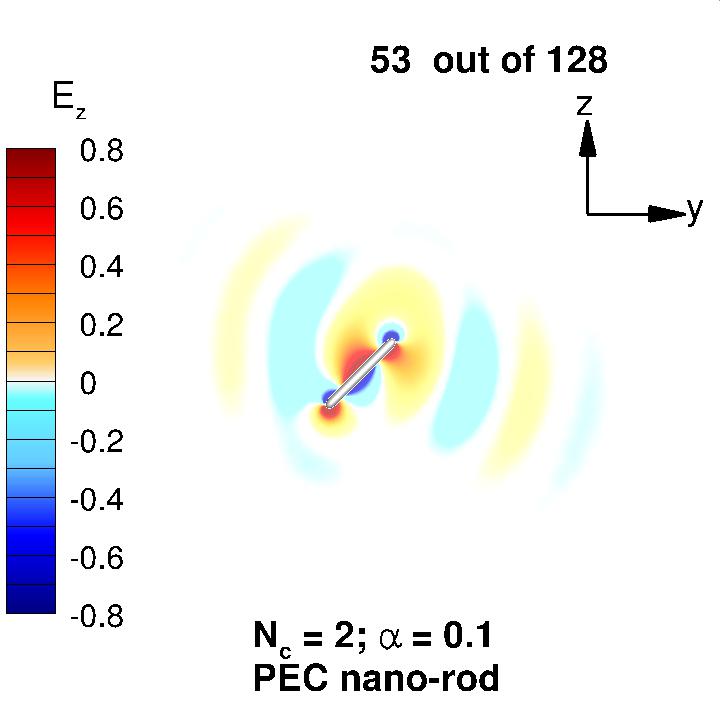} \\
    \includegraphics[width=.23\textwidth]{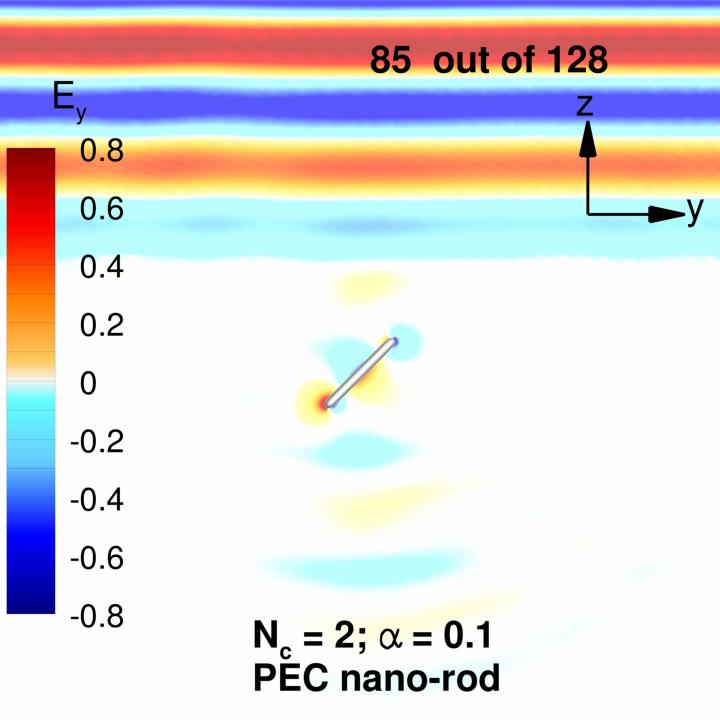} &
    \includegraphics[width=.23\textwidth]{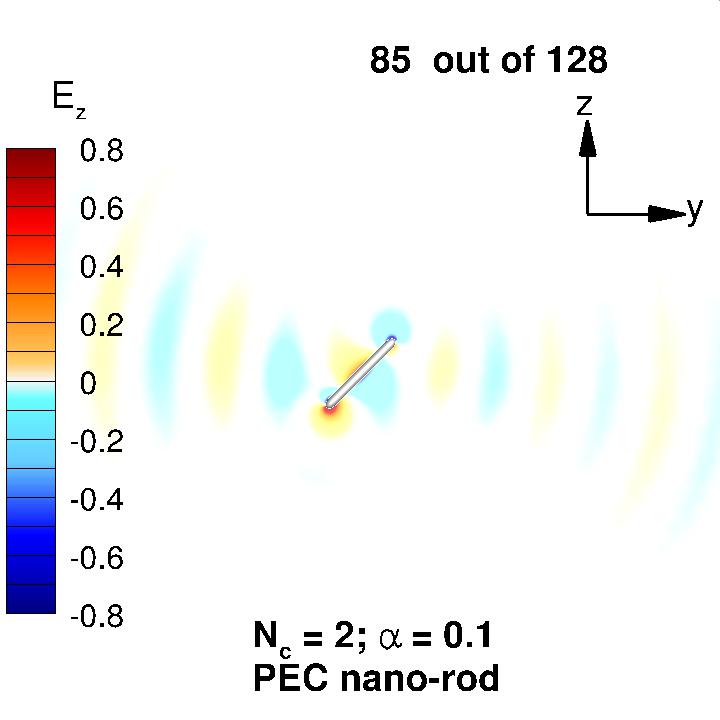} 
  \end{tabular}
  \caption{Snapshots of total amplitudes of $E_{y}$ and $E_{z}$ components in the $yz$-plane at different time points due to scattering of a plane wave pulse by a perfect electrical conducting nano-rod with length to width ratio 10. The incident wave is given by Eq.~\ref{eq:E_inc_y} with $k_0 L = 32\pi/20.1$ at the indicated time steps. The nano-rod is inclined at an angle of $45^\circ$ relative to the direction of pulse propagation along $z$-axis. See Visualization 5 in the Supplementary Material for full animation.}
\label{fig:PECrod}
\end{figure}

The scattering from an object with high aspect ratio is tackled next, both for a PEC and a dielectric object, a task which is rather challenging for conventional methods. 
In Fig.~\ref{fig:PECrod} and Fig.~\ref{fig:Dielecrod}, we compare the values of the $E_{y}$ and $E_{z}$ components of the total electrical field excited by the incident pulse given in Eq.~{\ref{eq:E_inc_y}} by a PEC and by a dielectic nano-rod with refractive index $n_{p}=3$ {\cite{Sun2017}, the surrounding medium has unit refractive index. 
The $y$-component of the electric field, $E_{y}$ is shown on the left and the $z$-component, $E_{z}$ on the right at three time instances in both figures. 
In both cases, the long axis of the nano-rod is oriented at an angle of $45^\circ$ to the propagation direction of the incident pulse. 
The surface of the axial symmetric nano-rod is defined by rotating an analytic closed curve about the long axis~\cite{Chwang1974}. 
The rod has length, $L$ and maximum cross-section diameter, $b$ at the middle of the rod, with $L/b = 10$. 
In both figures, the incident pulse is defined by Eq.~{\ref{eq:E_inc_y}} with $N_{c}=2$ and $\alpha = 0.1$. 
This corresponds to, $k_{0}L = 32\pi/20.1$ and $k_{0}b = 3.2\pi/20.1$ so that the length of the nano-rod is approximately the same wavelength, $\lambda_0 \equiv 2\pi/k_0$ of the incident pulse and the width of the nano-rod is about $\lambda_0/10$. 
As expected, the PEC nano-rod scatters much more strongly than its dielectric counterpart. 
This is evident by comparing the snapshots at the same time step in Fig.~\ref{fig:PECrod} and Fig.~\ref{fig:Dielecrod}. 
At frame 31, when nano-rod is near the center of the pulse, the scattering due to the PEC nano-rod is larger and consequently the perturbation of the total field is more severe. 
Towards the tail of the pulse at frame 53, the scattered field around the PEC is again much more pronounced than that of the dielectric nano-rod. After the pulse has passed, the scattered field from the dielectric nano-rod is essentially zero, whereas that due to the PEC is still creating a scattering effect which slowly dies out. 
This effect is even more obvious in the supplementary Visualizations 5 and 6). 
As a check of our numerical implementation, the scattering goes to zero smoothly, as expected, when the nano-rod is made transparent by taking the limit of its refractive index to be the same as that of the surrounding medium, $n_{p} \rightarrow 1$. 

\begin{figure}[t]   
\centering
  \begin{tabular}{@{}cc@{}}
    \includegraphics[width=.23\textwidth]{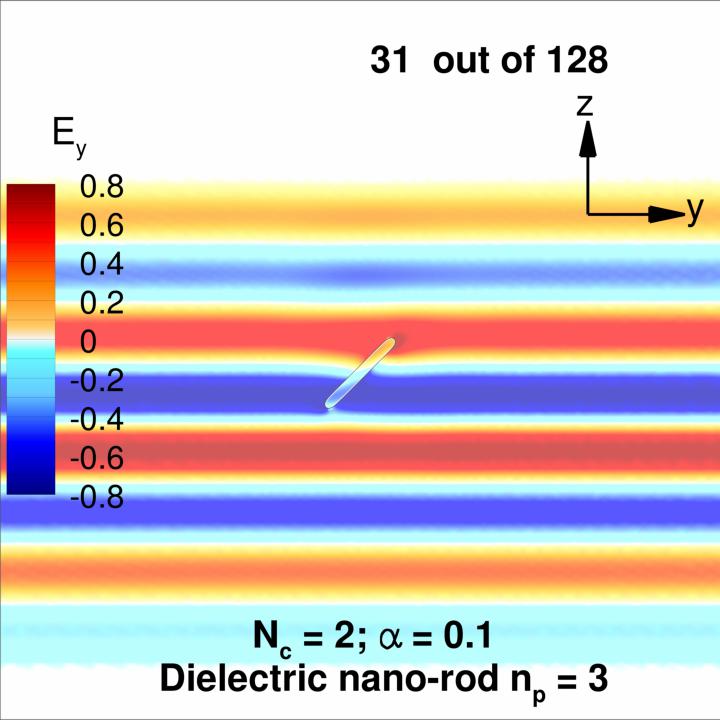} &
    \includegraphics[width=.23\textwidth]{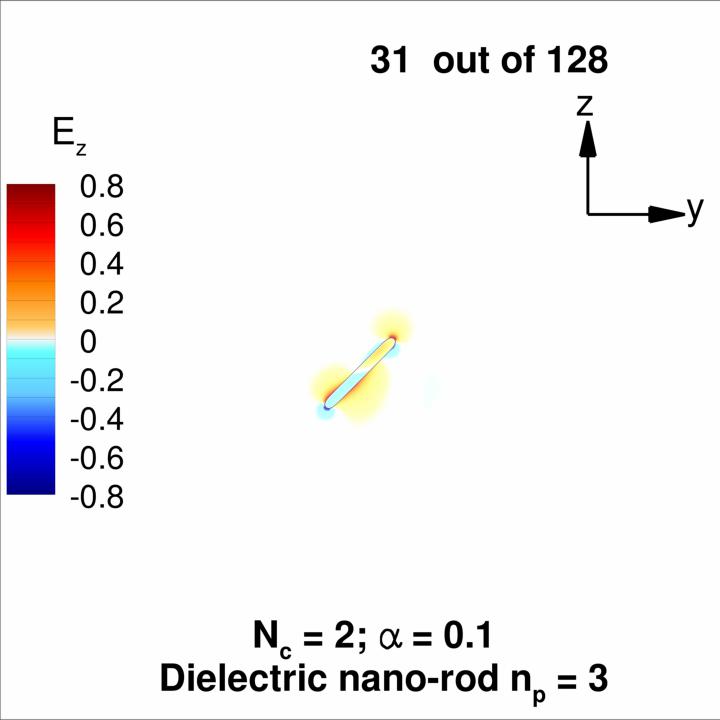} \\
    \includegraphics[width=.23\textwidth]{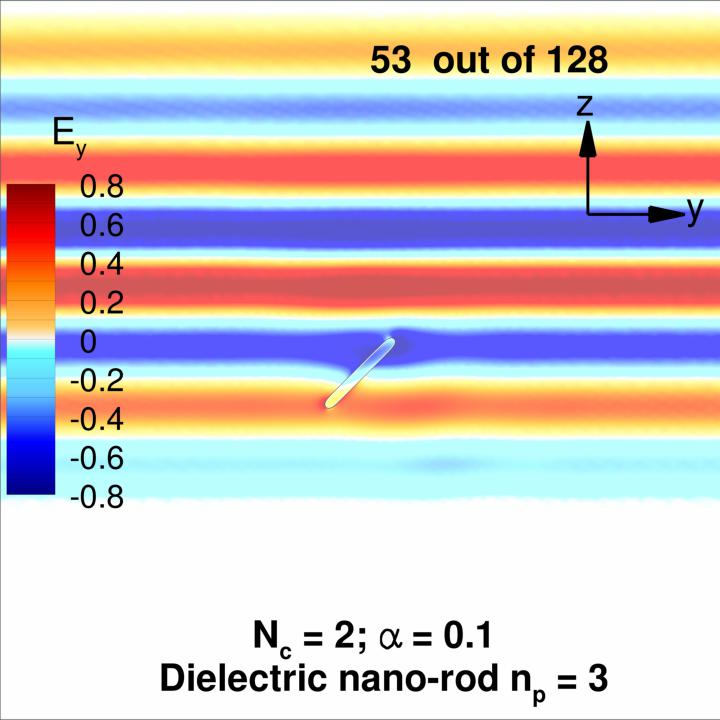} &
    \includegraphics[width=.23\textwidth]{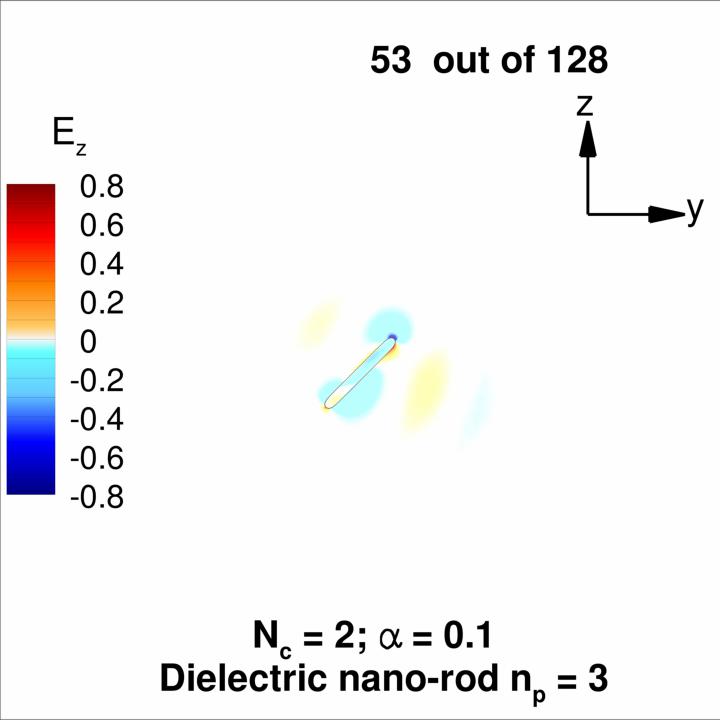} \\
    \includegraphics[width=.23\textwidth]{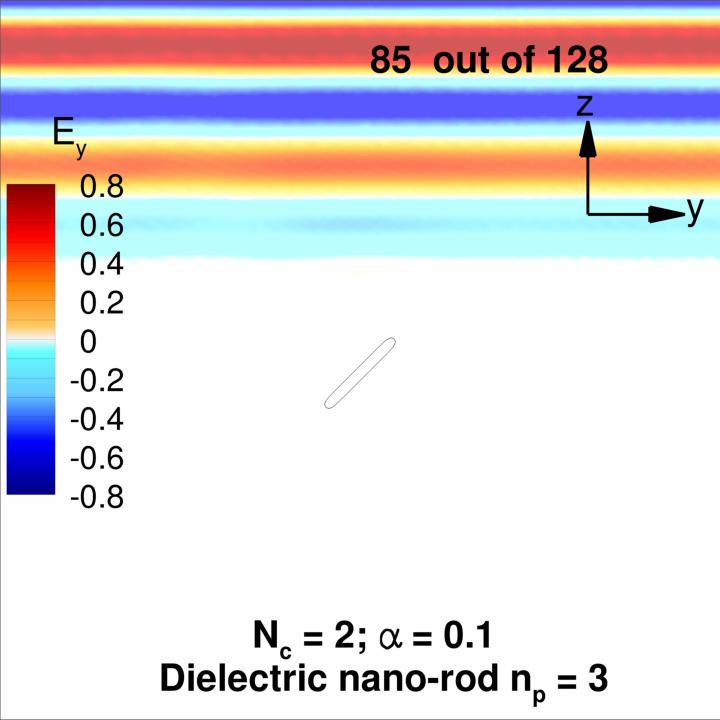} &
    \includegraphics[width=.23\textwidth]{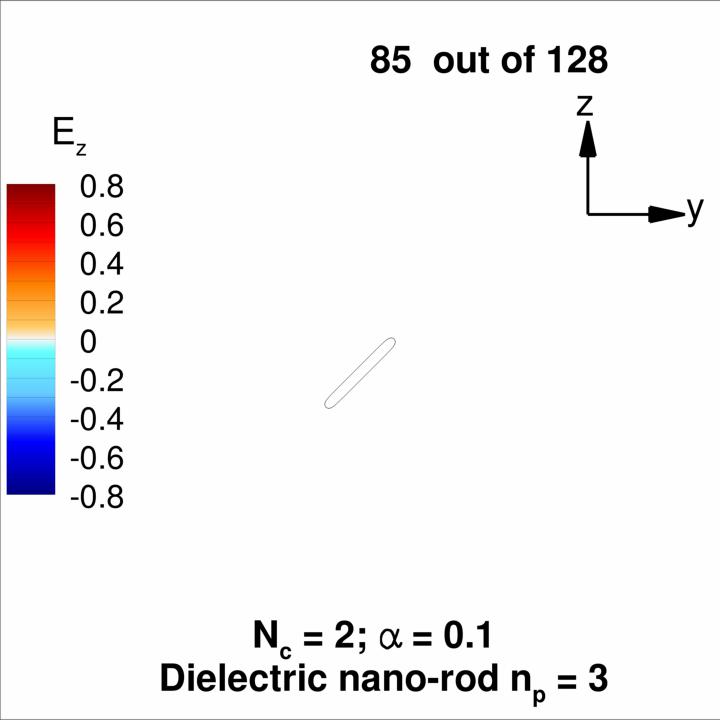}  
  \end{tabular}
  \caption{As Fig.~\ref{fig:PECrod}, but replacing the perfect electric conducting nano-rod by a dielectric nano-rod with refractive index $n_{p}=3$ relative to the surrounding medium. See Visualization 6 in the Supplementary Material for full animation.}
\label{fig:Dielecrod}
\end{figure}

\section{Conclusions}
We have leveraged on the direct nature and numerical stability of the field only formulation of the solution of Maxwell's equations in the frequency domain~\cite{Klaseboer2017a,Sun2017} to extend it to find time domain solutions using the Fourier transform. 
The field only formulation solves directly for the field components instead of first finding the surface current and then computing the field by post-processing. It is therefore particularly suitable for situations in which one is interested in the field amplitudes, phases and directions, especially near material boundaries - often desired in the design of microphotonic devices or in surface enhanced Raman spectroscopy.

In contrast, because of the singular nature of the surface integrals for the surface currents, the post processing step to find field components from the surface currents becomes unstable numerically as the field point becomes close to a boundary.
Although only results for the field in the $yz$-plane of symmetry are given for simpler visualization purposes, the complex 3D structure of the field around the scatterers are readily available. 

Although the finite difference time domain (FDTD) method also solves directly for the field components, it is necessary to discretise the 3D spatial domain as opposed to only to focus on 3D boundaries in surface integral formulations. This becomes especially beneficial for spatial regions with different characteristic length scales such as near the rim of the PEC bowl and around the sharp ends of the nano-rods in our examples. The 3D discretization problem can be challenging for conventional methods. Furthermore, when the natural orientation of the scatterer is at an angle to the direction of field propagation, care is needed in the construction of the 3D spatial mesh to avoid spurious dispersion effects~\cite{Taflove1988}. 

The present non-singular boundary integral formulation~\cite{Sun2015} for the solutions of the Helmholtz equation for the field components, is unaffected by numerical instabilities at the low frequency regime or in the regime when the characteristic length scale of the problem is small compare to the wavelength. In this low frequency or long wavelength regime, the conventional boundary integral equation approach will exhibit numerically troublesome near-singular behavior sometimes referred to as the zero frequency catastrophe that limits the precision that can be attained in calculations that involve first having to find induced surface currents~\cite{Vico2016}. In contrast, the present approach is numerically robust for wavelengths that span the electrostatic to approaching the geometric optics regimes without the need to modify the algorithm.

The absence of singularities in the integral equations also means numerical instabilities do not arise in multiple scattering effects due to scatterers in close proximity or in the evaluation of field values close to material boundaries. And in the absence of singular kernels, it is easy to use quadratic surface elements with consistent quadratic interpolation of the continuous integrands~\cite{Klaseboer2017a,Klaseboer2017b}, as we have done so in our numerical examples, to evaluate the surface integrals. This means higher order precision can be obtained with fewer number of nodes compared to using planar surface elements 

For a fixed set of scatterers, the boundary integral equations only need to be solved once from which the space-time variation of the total field for different incident pulses can be found by inverse Fourier transform.

In summary, the present approach involves solving non-singular scalar surface integral equations directly for the electric field components on the boundary of scatterers. In contrast, the conventional surface integral equation approach involves solving integral equations with singular kernels for surface currents. Although there are well developed methods to deal with such issues in the conventional surface integral approach, the integral equations become numerically ill-conditioned when the characteristic length scales of the scatterers are small compared to the wavelength. Since the electric field is obtained from the surface current by post processing, the presence of singularities in the integral equations means that the precision of field values at or near surfaces is compromised. The present approach avoids such issues by having to solve 4 scalar Helmholtz equations by a non-singular integral equation formulation. Thus, in handling the time evolution by Fourier transform, the robustness of our method at all frequencies is a distinct advantage.

Compared to the finite difference time domain (FDTD) method, the ability to achieve a reduction in dimension by only having to solve the problem on the boundary of scatterers also removes the challenge of having to discretize a complex  3D domain when there are vastly different characteristic length scales and spatial directions in a problem. In the FDTD approach, changing incident field for the same set of scatterers, requires repeating the solution process with each new incident field. In contrast, with the present approach, once the frequency domain problem has been found for a fixed set of scatterers, the solution can be used to find the response to different incident fields by simply taking the inverse Fourier transform.  


\bigskip
\noindent{\textbf{\textsf{Funding.}} This work is supported by the Australian Research Council (ARC) through a Discovery Project Grant (DP170100376) to DYCC and a Discovery Early Career Researcher Award (DE150100169) to QS.}

\bibliography{Main}

\bibliographyfullrefs{Main}
 
\end{document}